\documentclass[sigconf]{acmart}
\AtBeginDocument{%
  \providecommand\BibTeX{{%
    \normalfont B\kern-0.5em{\scshape i\kern-0.25em b}\kern-0.8em\TeX}}}

\setcopyright{acmcopyright}

\copyrightyear{2021}
\acmYear{2021}
\setcopyright{acmcopyright}\acmConference[WSDM '21]{Proceedings of the Fourteenth ACM International Conference on Web Search and Data Mining}{March 8--12, 2021}{Virtual Event, Israel}
\acmBooktitle{Proceedings of the Fourteenth ACM International Conference on Web Search and Data Mining (WSDM '21), March 8--12, 2021, Virtual Event, Israel}
\acmPrice{15.00}
\acmDOI{10.1145/3437963.3441800}
\acmISBN{978-1-4503-8297-7/21/03}

\newcommand{\ie}{\emph{i.e., }}
\newcommand{\eg}{\emph{e.g., }}

\newcommand{\wrt}{\emph{w.r.t. }}

\newcommand{\aka}{\emph{a.k.a. }}
\usepackage{subfigure}

\usepackage{algorithm}  
\usepackage{algorithmicx}  
\usepackage{algpseudocode}  
\usepackage{amsmath}  
\usepackage{enumitem}

\floatname{algorithm}{Algorithm}

\begin{document}
\fancyhead{}

%%
%% The "title" command has an optional parameter,
%% allowing the author to define a "short title" to be used in page headers.
%\title{Robust Training of Deep Learning based Recommenders \\ with Negative Implicit Interactions}
\title{Denoising Implicit Feedback for Recommendation}

%%
%% The "author" command and its associated commands are used to define
%% the authors and their affiliations.
%% Of note is the shared affiliation of the first two authors, and the
%% "authornote" and "authornotemark" commands
%% used to denote shared contribution to the research.

\author{Wenjie Wang}
\email{wenjiewang96@gmail.com}
\affiliation{%
  \institution{National University of Singapore}
}
\author{Fuli Feng}
\authornote{Corresponding author: Fuli Feng (fulifeng93@gmail.com).}
\email{fulifeng93@gmail.com}
\affiliation{%
  \institution{National University of Singapore}
}

\author{Xiangnan He}
\email{hexn@ustc.edu.cn}
\affiliation{%
  \institution{University of Science and Technology of China}
}
\author{Liqiang Nie}
\email{nieliqiang@gmail.com}
\affiliation{%
  \institution{Shandong University}
}

\author{Tat-Seng Chua}
\email{dcscts@nus.edu.sg}
\affiliation{%
  \institution{National University of Singapore}
}
% \thanks{* Corresponding author: Fuli Feng (fulifeng93@gmail.com)}

\begin{abstract}
The ubiquity of implicit feedback makes them the default choice to build online recommender systems. While the large volume of implicit feedback alleviates the data sparsity issue, the downside is that they are not as clean in reflecting the actual satisfaction of users. For example, in E-commerce, a large portion of clicks do not translate to purchases, and many purchases end up with negative reviews. As such, it is of critical importance to account for the inevitable noises in implicit feedback for recommender training. However, little work on recommendation has taken the noisy nature of implicit feedback into consideration. 

In this work, we explore the central theme of denoising implicit feedback for recommender training. 
We find serious negative impacts of noisy implicit feedback, \ie fitting the noisy data hinders the recommender from learning the actual user preference. Our target is to identify and prune the noisy interactions, so as to improve the efficacy of recommender training. 
By observing the process of normal recommender training, we find that noisy feedback typically has large loss values in the early stages. Inspired by this observation, we propose a new training strategy named \textit{Adaptive Denoising Training} (ADT), which adaptively prunes noisy interactions during training. 
Specifically, we devise two paradigms for adaptive loss formulation: \textit{Truncated Loss} that discards the large-loss samples with a dynamic threshold in each iteration; and \textit{Reweighted Loss} that adaptively lowers the weights of large-loss samples.
We instantiate the two paradigms on the widely used binary cross-entropy loss and test the proposed ADT strategies on three representative recommenders. Extensive experiments on three benchmarks demonstrate that ADT significantly improves the quality of recommendation over normal training.
\end{abstract}

%%
%% The code below is generated by the tool at http://dl.acm.org/ccs.cfm.
%% Please copy and paste the code instead of the example below.
%%


%%
%% Keywords. The author(s) should pick words that accurately describe
%% the work being presented. Separate the keywords with commas.

%%
%% This command processes the author and affiliation and title
%% information and builds the first part of the formatted document.
\maketitle

\section{Introduction}\label{sec:intro}

Recommender systems have been a promising solution for mining user preference over items in various online services such as E-commerce~\cite{nie2019multimodal}, news portals~\cite{Lu2018Between} and social media~\cite{ren2017social, wang2018chat}. As the clue to user choices, implicit feedback (\eg click and purchase) are typically the default choice to train a recommender due to their large volume. However, prior work~\cite{Hu2008CF, Lu2018Between, Wen2019Leveraging} points out the gap between implicit feedback and the actual user satisfaction due to the prevailing presence of \textit{noisy interactions} (\aka \textit{false-positive interactions}) where the users dislike the interacted item. For instance, in E-commerce, a large portion of purchases end up with negative reviews or being returned. This is because implicit interactions are easily affected by the first impression of users and other factors \cite{chen2020bias, wang2020click} such as caption bias \cite{Hofmann2012on} and position bias \cite{Jagerman2019To}. Moreover, existing studies~\cite{ Wen2019Leveraging} have demonstrated the detrimental effect of such false-positive interactions on user experience of online services. Nevertheless, little work on recommendation has taken the noisy nature of implicit feedback into consideration. 

In this work, we argue that such false-positive interactions would hinder a recommender from learning the actual user preference, leading to low-quality recommendations. Table~\ref{table1} provides empirical evidence on the negative effects of false-positive interactions when we train a competitive recommender, Neural Matrix Factorization (NeuMF)~\cite{He2017Neural}, on two real-world datasets. We construct a ``clean'' testing set by removing the false-positive interactions for recommender evaluation\footnote{Each false-positive interaction is identified by auxiliary information of post-interaction behaviors, \eg rating score ([1, 5]) $ < 3$, indicating that the interacted item dissatisfies the user. Refer to Section \ref{sec:study} for more details.}. As can be seen, training NeuMF with false-positive interactions (\ie \textit{normal training}) results in an average performance drop of 15.69\% and 6.37\% over the two datasets \wrt Recall@20 and NDCG@20, as compared to the NeuMF trained without false-positive interactions (\ie \textit{clean training}). As such, it is critical to account for the inevitable noises in implicit feedback and perform denoising.

\begin{figure}[t]
\vspace{0cm}
\setlength{\abovecaptionskip}{0.05cm}
\setlength{\belowcaptionskip}{-0.5cm}
\includegraphics[scale=0.58]{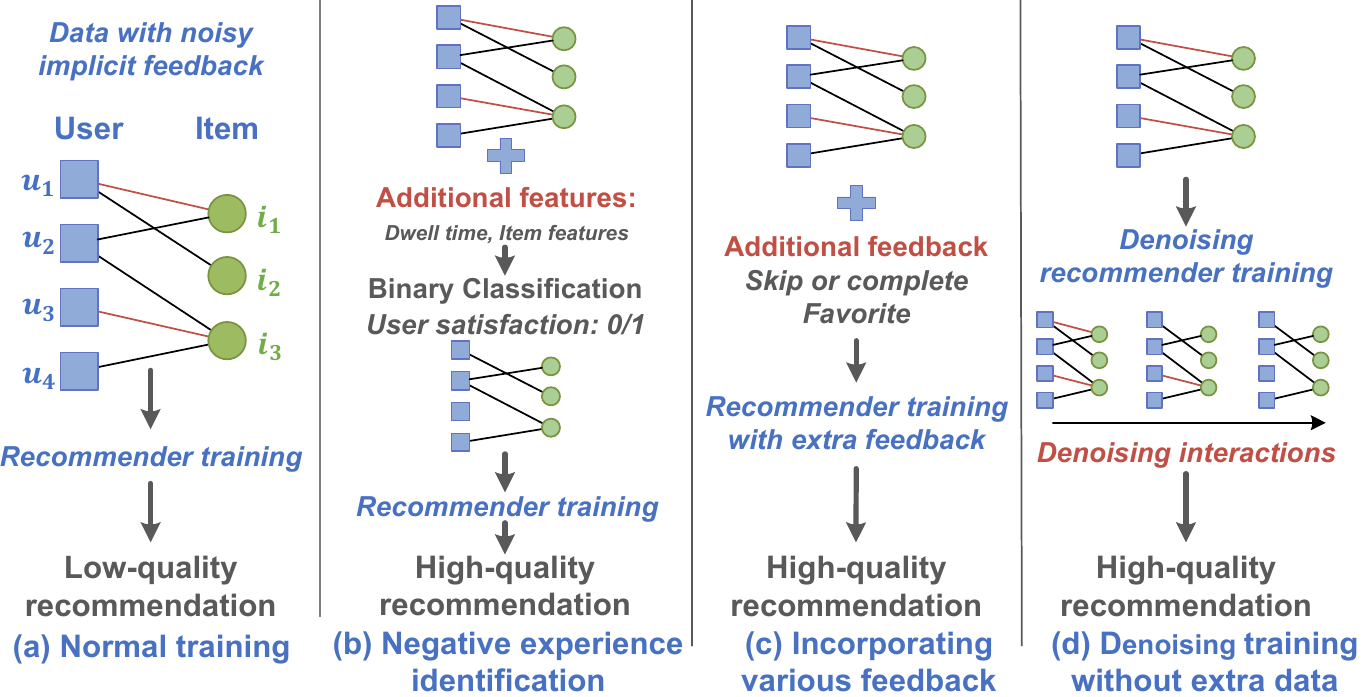}
\caption{The comparison between normal training (a); two prior solutions (b) and (c); and denoising training without extra data (d). Red lines in the user-item graph denote false-positive interactions.}
\label{Figure1}
\end{figure}

Indeed, some efforts \cite{fox2005Evaluating, Kim2014Modeling, Yang2012Exploiting} have been dedicated to eliminating the effects of false-positive interactions by: 1) negative experience identification~\cite{Kim2014Modeling} (illustrated in Figure~\ref{Figure1}(b)); and 2) the incorporation of various feedback \cite{Yang2012Exploiting, Yi2014Beyond} (shown in Figure~\ref{Figure1}(c)). The former processes the implicit feedback in advance by predicting the false-positive ones with additional user behaviors (\eg dwell time and gaze pattern) and auxiliary item features (\eg length of the item description)~\cite{Lu2018Between}. The latter incorporates extra feedback (\eg favorite and skip) into recommender training to prune the effects of false-positive interactions~\cite{Yi2014Beyond}. A key limitation with these methods is that they require additional data to perform denoising, which may not be easy to collect.  Moreover, extra feedback (\eg rating and favorite) is often of a smaller scale, which will suffer from the sparsity issue. For instance, many users do not give any feedback after watching a movie or purchasing a product~\cite{Hu2008CF}.

%******************** Figure 1 *****************%
%
%
%******************* Figure 1 ***********************
%

This work explores denoising implicit feedback for recommender training, which automatically reduces the influence of false-positive interactions without using any additional data (Figure~\ref{Figure1}(d)). That is, we only count on the implicit interactions and distill signals of false-positive interactions across different users and items. Prior study on robust learning~\cite{lu2018MentorNet, han2018co} and curriculum learning~\cite{Bengio2009Curriculum} demonstrate that noisy interactions are relatively harder to fit into models, indicating distinct patterns of noisy interactions' loss values in the training procedure. Primary experiments across different recommenders and datasets (\eg Figure~\ref{fig:observation}) reveal similar phenomenon: the loss values of false-positive interactions are larger than those of the true-positive ones in the early stages of training. Consequently, due to the larger loss, false-positive interactions can largely mislead the recommender training in the early stages. Worse still, the recommender ultimately fits the false-positive interactions due to its high representation capacity, which could be overfitting and hurt the generalization. As such, a potential idea of denoising is to reduce the impact of false-positive interactions, \eg pruning the interactions with large loss values, where the key challenge is to simultaneously decrease the sacrifice of true-positive interactions.

To this end, we propose Adaptive Denoising Training (ADT) strategies for recommenders, which dynamically prunes the large-loss interactions along the training process. To avoid the lost of generality, we focus only on formulating the training loss, which can be applied to any differentiable models. In detail, we devise two paradigms to formulate the training loss: 1) \textit{Truncated Loss}, which discards the large-loss interactions dynamically, and 2) \textit{Reweighted Loss}, which adaptively reweighs the interactions. For each training iteration, the Truncated Loss removes the hard interactions (\ie large-loss ones) with a dynamic threshold which is automatically updated during training. The Reweighted Loss dynamically assigns hard interactions with smaller weights to weaken their effects on the optimization. We instantiate the two functions on the basis of the widely used binary cross-entropy loss. On three benchmarks, we test the Truncated Loss and Reweighted Loss over three representative recommenders: Generalized Matrix Factorization (GMF)~\cite{He2017Neural}, NeuMF~\cite{He2017Neural}, and Collaborative Denoising Auto-Encoder (CDAE)~\cite{wu2016collaborative}. The results show significant performance improvements of ADT over normal training. 
Codes and data are publicly available\footnote{\url{https://github.com/WenjieWWJ/DenoisingRec.}}.

Our main contributions are summarized as:
\begin{itemize}[leftmargin=*]
    \item We formulate the task of denoising implicit feedback for recommender training. We find the negative effect of false-positive interactions and identify their large-loss characteristics.
    \item We propose Adaptive Denoising Training to prune the large-loss interactions dynamically, which introduces two paradigms to formulate the training loss: Truncated Loss and Reweighted Loss.
    \item We instantiate two paradigms on the binary cross-entropy loss and apply ADT to three representative recommenders. Extensive experiments on three benchmarks validate the effectiveness of ADT in improving the recommendation quality.
\end{itemize}

%
%************************ table 1**************%
%
\begin{table}[]
\setlength{\abovecaptionskip}{0cm}
\setlength{\belowcaptionskip}{0cm}
\caption{Results of the clean training and normal training over NeuMF. \#Drop denotes the relative performance drop of normal training as compared to clean training.}
\label{table1}
\centering
\resizebox{0.48\textwidth}{!}{
\begin{tabular}{l|cc|cc}
\hline
\textbf{Dataset}     & \multicolumn{2}{c|}{\textbf{Adressa}} & \multicolumn{2}{c}{\textbf{Amazon-book}} \\ % \hline
\textbf{Metric}      & \textbf{Recall@20}    & \textbf{NDCG@20}   & \textbf{Recall@20}   & \textbf{NDCG@20}   \\ \hline \hline
\textbf{Clean training} & 0.4040                & 0.1963             & 0.0293               & 0.0159             \\ %\hline
\textbf{Normal training}  & 0.3159                & 0.1886             & 0.0265               & 0.0145             \\ \hline
\textbf{\#Drop}      & 21.81\%               & 3.92\%            & 9.56\%               & 8.81\%             \\ \hline
\end{tabular}
%}
}
\vspace{-0.5cm}
\end{table}
%
%************************ table 1 **************%
%

\vspace{-0.15cm}
\section{Study on False-Positive Feedback}\label{sec:study}

The effect of noisy training samples \cite{brodley1999} has been studied in conventional machine learning tasks such as image classification~\cite{lu2018MentorNet, han2018co}. However, little attention has been paid to such effect on recommendation, which is inherently different from conventional tasks due to the relations across training samples, \eg interactions on the same item. We investigate the effects of false-positive interactions on recommender training by comparing the performance of recommenders trained with all observed user-item interactions including the false-positive ones (normal training); and without false-positive interactions (clean training). An interaction is identified as false-positive or true-positive one according to the explicit feedback. For instance, a purchase is false-positive if the following rating score ([1, 5]) < 3. Although the size of such explicit feedback is typically insufficient for building robust recommenders in real-world scenarios, the scale is sufficient for a pilot experiment. 
% In detail, we train a competitive recommender model NeuMF under two different settings: 1) ``clean training'' which trains NeuMF on the true-positive interactions only; and 2) ``normal training'' which trains NeuMF on all observed user-item interactions. 
We evaluate the recommendation performance on a holdout clean testing set with only true-positive interactions kept, \ie the evaluation focuses on recommending more satisfying items to users. More details can be seen in Section~\ref{sec:protocol}.

\vspace{-0.15cm}
\paragraph{Results}

Table \ref{table1} summarizes the performance of NeuMF under normal training and clean training \wrt Recall@20 and NDCG@20 on two representative datasets, Adressa and Amazon-book. From Table \ref{table1}, we can observe that, as compared to clean training, the performance of normal training drops significantly (\eg 21.81\% and 9.56\% \wrt Recall@20 on Adressa and Amazon-book). This result shows the \textit{negative effects} of false-positive interactions on recommending satisfying items to users. Worse still, recommenders under normal training have higher risk to produce more false-positive interactions, which further hurt the user experience~\cite{Lu2018Between}. Despite the success of clean training in the pilot study, it is not a reasonable choice in practical applications because of the sparsity issues of reliable feedback such as rating scores. As such, it is worth exploring denoising implicit feedback such as click, view, or buy for recommender training.

\section{Method}
In this section, we detail the proposed Adaptive Denoising Training strategy for recommenders. Prior to that, task formulation and observations that inspire the strategy design are introduced. 

\vspace{-0.2cm}
\subsection{\textbf{Task Formulation}}

Generally, the target of recommender training is to learn user preference from user feedback, \ie learning a scoring function $\hat{y}_{ui} = f(u, i | \Theta)$ to assess the preference of user $u$ over item $i$ with parameters $\Theta$. Ideally, the setting of recommender training is to learn $\Theta$ from a set of reliable feedback between $N$ users ($\mathcal{U}$) and $M$ items ($\mathcal{I}$). That is, given $\mathcal{D}^{\ast} = \{(u, i, y^{\ast}_{ui}) | u \in \mathcal{U}, i \in \mathcal{I}\}$, we learn the parameters $\Theta^{\ast}$ by minimizing a recommendation loss over $\mathcal{D}^{\ast}$, \eg the binary Cross-Entropy (CE) loss:
%
%************************ formula 4 **************%
%
\begin{equation}\small\notag
\begin{aligned}%\small
& \mathcal{L}_{CE}\left(\mathcal{D}^{\ast}\right) = - \sum_{(u, i,  y^{\ast}_{ui}) \in \mathcal{D}^{\ast}} y^{\ast}_{ui}\log\left(\hat{y}_{ui}\right) + \left(1 - y^{\ast}_{ui}\right)\log\left(1-\hat{y}_{ui}\right), 
\end{aligned}
\end{equation}
%
%************************ formula 4 **************%
%
where $y^{\ast}_{ui} \in \{0, 1\}$ represents whether the user $u$ really prefers the item $i$. The recommender with $\Theta^{\ast}$ would be reliable to generate high-quality recommendations. In practice, due to the lack of large-scale reliable feedback, recommender training is typically formalized as: $\bar{\Theta} = \min \mathcal{L}_{CE}(\mathcal{\bar{D}})$, where $\mathcal{\bar{D}} = \{(u, i, \bar{y}_{ui}) | u \in \mathcal{U}, i \in \mathcal{I}\}$ is a set of implicit interactions. $\bar{y}_{ui}$ denotes whether the user $u$ has interacted with the item $i$, such as click and purchase.

\begin{figure}[htb]
\vspace{-0.2cm}
\setlength{\abovecaptionskip}{0cm}
\setlength{\belowcaptionskip}{-0.4cm}
\vspace{-0.2cm}
\includegraphics[scale=0.35]{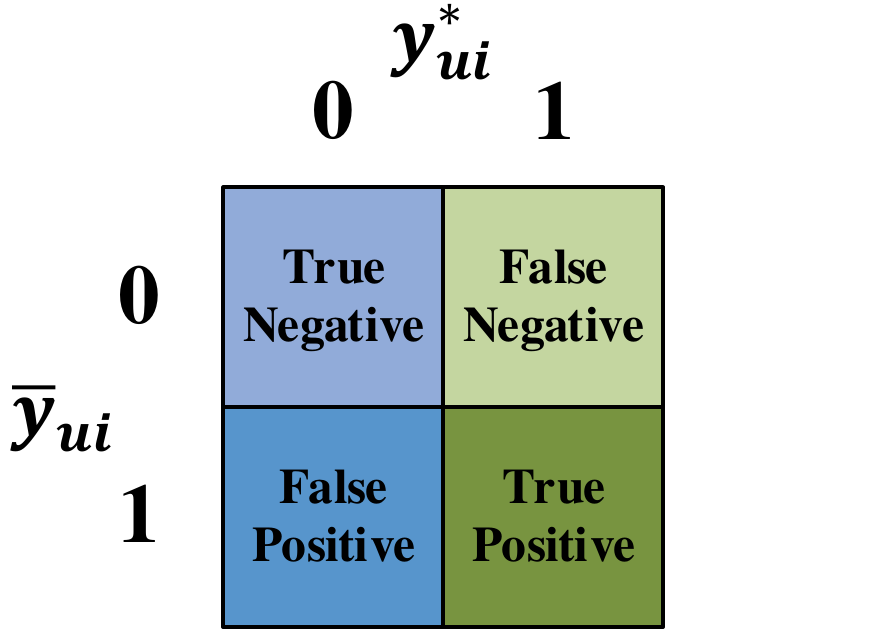}
\caption{Four types of implicit interactions.}
\label{fig:grid}
\end{figure}

However, due to the existence of noisy interactions which would mislead the learning of user preference, the typical recommender training might result in a poor model (\ie $\bar{\Theta}$) that lacks generalization ability on the clean testing set. As such, we formulate a \textit{denoising recommender training} task as:
\begin{equation}\small
\begin{aligned}
& {\Theta}^{\ast} = \min \mathcal{L}_{CE}\big(denoise(\mathcal{\bar{D}})\big),
\end{aligned}
\end{equation}
aiming to learn a reliable recommender with parameters $\Theta^{\ast}$ by denoising implicit feedback, \ie pruning the impact of noisy interactions. Formally, by assuming the existence of inconsistency between $y^{\ast}_{ui}$  and $\bar{y}_{ui}$, we define noisy interactions as $\left\{ (u, i) | y^{\ast}_{ui}=0 \land \bar{y}_{ui}=1 \right\}$. According to the value of $y^{\ast}_{ui}$ and $\bar{y}_{ui}$, we can separate implicit feedback into four categories similar to a confusion matrix as shown in Figure~\ref{fig:grid}.
%
%************************ Figure 3 **************%
%
%
%************************ Figure 3 **************%
%

In this work, we focus on denoising false-positive interactions and omit the false-negative ones. This is because positive interactions are sparse and false-positive interactions can thus be more influential than false-negative ones. Note that we do not incorporate any additional data such as explicit feedback into the task of denoising recommender training. This is because such feedback is of a smaller scale in most cases, suffering more severely from the sparsity issue.

%
%************************ Figure 2 **************%
%

\begin{figure}[h]
\vspace{-0.2cm}
\setlength{\abovecaptionskip}{-0.10cm}
\setlength{\belowcaptionskip}{-0.5cm}
  \centering 
  \hspace{-0.2in}
  \subfigure[Whole training process]{
    % \vspace{-0.1in}
    \label{fig2:subfig:a}
    \includegraphics[width=1.5in]{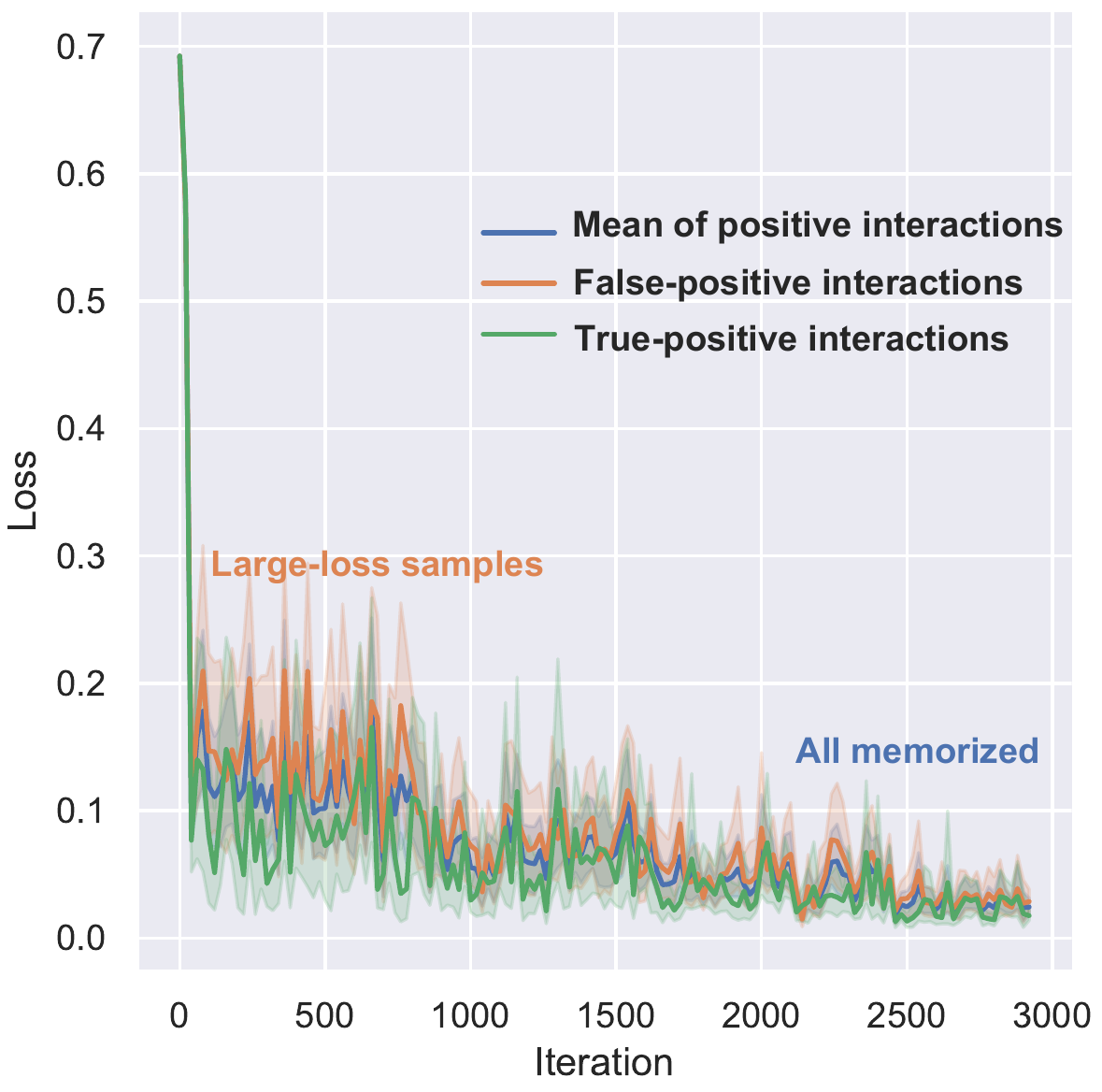}} 
%   \hspace{-0.4in}
  \subfigure[Early training stages]{ 
    % \vspace{-0.1in}
    \label{fig2:subfig:b} 
    \includegraphics[width=1.5in]{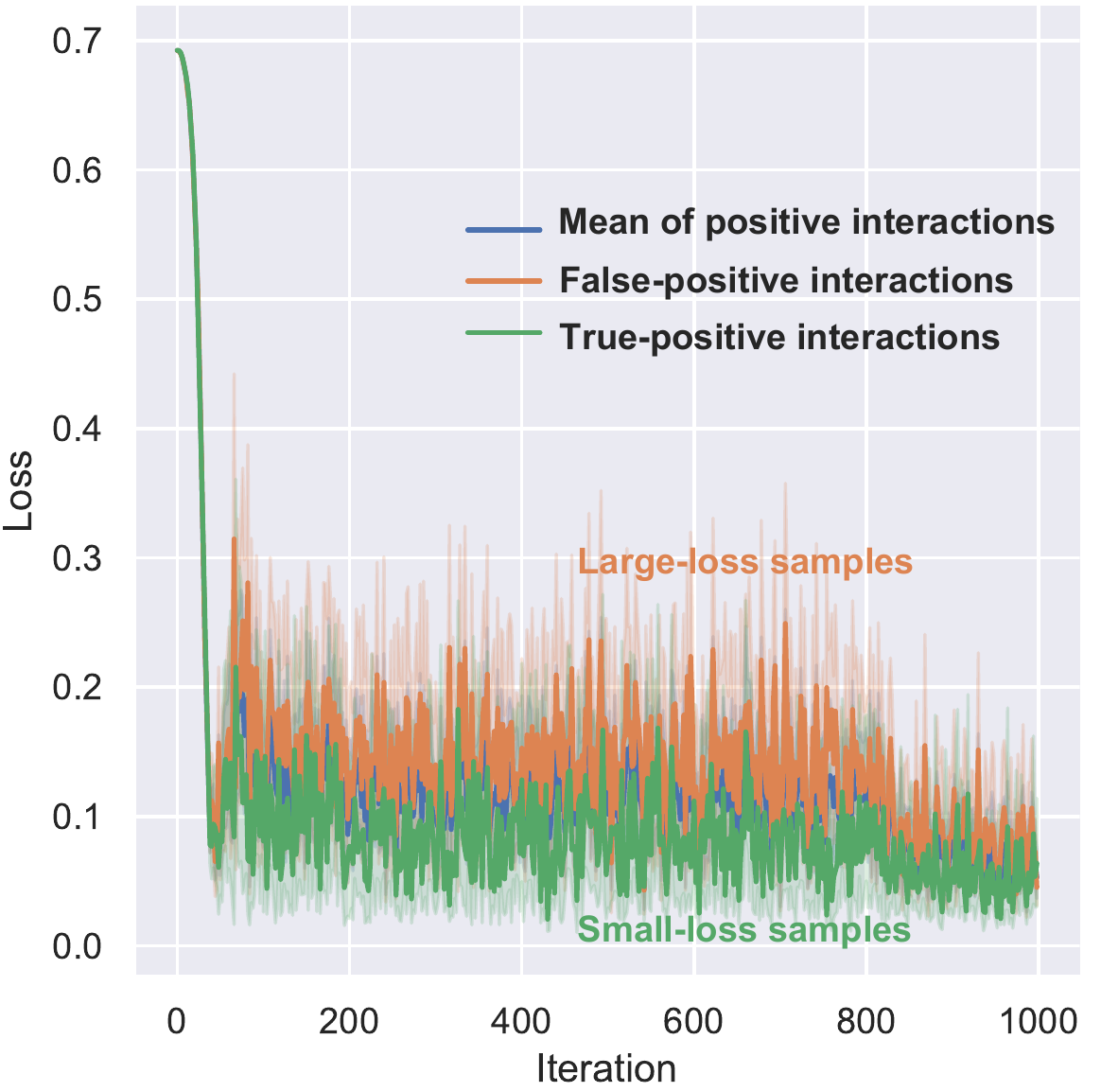}} 
%   \hspace{-0.5in} 
%   \vspace{-0.1in}
  \caption{The training loss of true- and false-positive interactions on Adressa in the normal training of NeuMF.} 
  \label{fig:observation}
\end{figure}
%
%************************ Figure 2 **************%
%

\subsection{Observations}\label{sec:observations}

\paragraph{{False-positive interactions are harder to fit in the early stages.}}
According to the theory of robust learning~\cite{lu2018MentorNet, han2018co} and curriculum learning~\cite{Bengio2009Curriculum}, easy samples are more likely to be the clean ones and fitting the hard samples may hurt the generalization. To verify its existence in recommendation, we compare the loss of true- and false-positive interactions along the process of normal training. Figure \ref{fig:observation} shows the results of NeuMF on the Adressa dataset. 
Similar trends are also found over other recommenders and datasets (see more details in Section \ref{sec:memorization}). 
From Figure \ref{fig:observation}, we observe that:
\begin{itemize}[leftmargin=*]
    \item Ultimately, the loss of both of true- and false-positive interactions converges to a stable state with close values, which implies that NeuMF fits both of them well. It reflects that deep recommender models can ``memorize'' all the training interactions, including the noisy ones. As such, if the data is noisy, the memorization will lead to poor generalization performance. 
    \item The loss values of true- and false-positive interactions decrease differently in the early stages of training. From Figure \ref{fig2:subfig:b}, we can see that the loss of false-positive interactions is clearly larger than that of the true-positive ones, which indicates that false-positive interactions are harder to memorize than the true-positive ones in the early stages. The reason might be that false-positive ones represent the items users dislike, and they are more similar to the items the user didn't interact with (\ie the negative interactions). The findings also support the prior theory in robust learning and curriculum learning~\cite{han2018co, Bengio2009Curriculum}.
\end{itemize}
Overall, the results are consistent with the memorization effect~\cite{arpit2017closer}: deep recommender models will learn the easy and clean patterns in the early stage, and eventually memorize all training interactions~\cite{han2018co}.

%
%************************ Figure 4 **************%
%

\begin{figure}
\vspace{0cm}
\setlength{\abovecaptionskip}{0cm}
\setlength{\belowcaptionskip}{-0.5cm}
\includegraphics[scale=0.4]{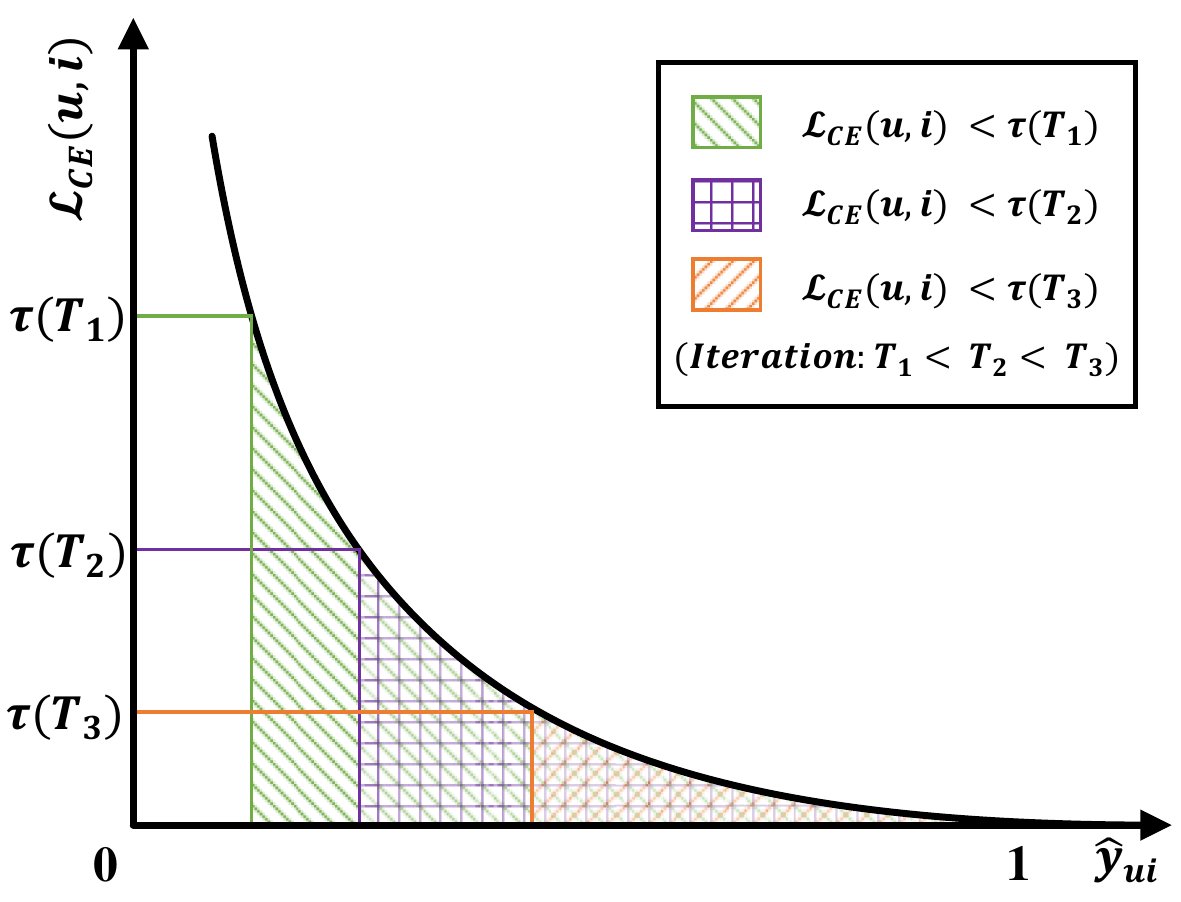}
\caption{Illustration of T-CE loss for the observed interactions (\ie $\bar{y}_{ui} = 1$). $T_i$ is the iteration number and $\tau(T_i)$ refers to the threshold. Dash area indicates the effective loss and the loss values larger than $\tau(T_i)$ are truncated.}
\label{fig:truncatedloss}
\end{figure}
%
%************************ Figure 4 **************%
%

\subsection{Adaptive Denoising Training}

According to the observations, we propose ADT strategies for recommenders, which estimate $P(y^{\ast}_{ui}=0 | \bar{y}_{ui}=1, u, i)$ according to the training loss. To reduce the impact of false-positive interactions, ADT dynamically prunes the hard interactions (\ie large-loss ones) during training. In particular, ADT either \textit{discards} or \textit{reweighs} the interactions with large loss values to reduce their influences on the training objective. Towards this end, we devise two paradigms to formulate loss functions for denoising training:
\begin{itemize}[leftmargin=*]
\item \textit{Truncated Loss.} This is to truncate the loss values of hard interactions to 0 with a dynamic threshold function.
\item \textit{Reweighted Loss.} It adaptively assigns hard interactions with smaller weights during training.
\end{itemize}
Note that the two paradigms formulate various recommendation loss functions, \eg CE loss, square loss~\cite{Rosasco2004are}, and BPR loss~\cite{rendle2009bpr}. In the work, we take CE loss as an example to elaborate them.

\subsubsection{\textbf{Truncated Cross-Entropy Loss}}

Functionally speaking, the Truncated Cross-Entropy (shorted as T-CE) loss discards positive interactions according to the values of CE loss. Formally, we define it as:
%
%************************ formula 5 **************%
%
\begin{equation}\small
\mathcal{L}_{\textit{T-CE}}(u, i) = 
\begin{cases}
0, & \mathcal{L}_{CE}(u, i) > \tau \land \bar{y}_{ui}=1 \\
\mathcal{L}_{CE}(u, i), & \text{otherwise,}
\end{cases}
\end{equation}
%
%************************ formula 5 **************%
%
where $\tau$ is a pre-defined threshold. The T-CE loss removes any positive interactions with CE loss larger than $\tau$ from the training. While this simple T-CE loss is easy to interpret and implement, the fixed threshold may not work properly. This is because the loss value is decreasing with the increase of training iterations. Inspired by the dynamic adjustment ideas~\cite{kingma2014adam, feng2018learning}, we replace the fixed threshold with a dynamic threshold function $\tau(T)$ \wrt the training iteration $T$, which changes the threshold value along the training process (Figure \ref{fig:truncatedloss}).
Besides, since the loss values vary across different datasets, it would be more flexible to devise $\tau(T)$ as a function of the drop rate $\epsilon(T)$. There is a bijection between the drop rate and the threshold, \ie we can calculate the threshold to filter out interactions in a training iteration according to the drop rate $\epsilon(T)$.

Based on prior observations, a proper drop rate function should have the following properties: 
1) $\epsilon(\cdot)$ should have an upper bound to limit the proportion of discarded interactions so as to prevent data missing;
2) $\epsilon(0) = 0$, \ie it should allow all the interactions to be fed into the models in the beginning; 
and 3) $\epsilon(\cdot)$ should increase smoothly from zero to its upper bound, so that the model can learn and distinguish the true- and false-positive interactions gradually.

Towards this end, we formulate the drop rate function as:
%
%************************ formula 5 **************%
%
\begin{equation}\small
\epsilon(T) = min(\alpha T, \epsilon_{max}),
\end{equation}
%
%************************ formula 5 **************%
%
where $\epsilon_{max}$ is an upper bound and $\alpha$ is a hyper-parameter to adjust the pace to reach the maximum drop rate. Note that we increase the drop rate in a linear fashion rather than a more complex function such as a polynomial function or a logarithm function. Despite the expressiveness of these functions, they will inevitably increase the number of hyper-parameters, resulting in the increasing cost of tuning a recommender. The whole algorithm is explained in Algorithm \ref{algo:T-CE}. As discarding the hard interactions, T-CE loss is symmetrically contrary to the Hinge loss \cite{Goodfellow2016deep}. And thus T-CE loss prevents the model from overfitting hard interactions for denoising.

\begin{algorithm}[t]
	\caption{Adaptive Denoising Training with T-CE loss}  
	\label{algo:T-CE}
	\begin{algorithmic}[1]
		\Require trainable parameters $\Theta$, training interactions $\mathcal{\bar{D}}$, the iterations $T_{max}$, learning rate $\eta$, $\epsilon_{max}$, $\alpha$, $\mathcal{L}_{CE}$, parameter optimization function $\nabla$.
		\For{$T = 1 \to T_{max}$} \algorithmiccomment{shuffle interactions every epoch}
		\State \textbf{Fetch} mini-batch data $\mathcal{\bar{D}}_{pos}$ from $\mathcal{\bar{D}}$
		\State \textbf{Sample} unobserved interactions $\mathcal{\bar{D}}_{neg}$
		\State \textbf{Define} $\mathcal{\bar{D}}_{T} = \mathcal{\bar{D}}_{pos} \cup \mathcal{\bar{D}}_{neg}$
		\State \textbf{Obtain} $\mathcal{\hat{D}} = \mathop{\arg\max}\limits_{\mathcal{\hat{D}}\subset \mathcal{\bar{D}}_{pos}, |\mathcal{\hat{D}}| = \lfloor \epsilon(T)*|\mathcal{\bar{D}}_{\textit{T}}|\rfloor} \sum_{(u,i) \in \mathcal{\hat{D}}} \mathcal{L}_{CE}(u, i | \Theta_{\textit{T-1}})$
		\State \textbf{Obtain} $\mathcal{{D}^*} = \mathcal{\bar{D}}_{T} - \mathcal{\hat{D}}$
		\State \textbf{Update} $\Theta_{\textit{T}} = \nabla(\Theta_{\textit{T-1}}, \eta, \mathcal{L}_{CE}, \mathcal{{D}^*})$
% 		\State \textbf{Update} $\Theta_{\textit{T}} = \Theta_{\textit{T-1}} - \eta\nabla\frac{1}{|\mathcal{\hat{D}}|}\sum_{u,i \in \mathcal{\hat{D}}} \mathcal{L}_{CE}(u, i | \Theta_{\textit{T-1}})$
		\State \textbf{Update} $\epsilon(T) = min(\alpha T, \epsilon_{max})$
		\EndFor 
		\Ensure the optimized parameters $\Theta_{T_{max}}$ of the recommender
	\end{algorithmic}
\end{algorithm}
\setlength{\textfloatsep}{0.1cm}

\subsubsection{\textbf{Reweighted Cross-Entropy Loss}}
Functionally speaking, the Reweighted Cross-Entropy (shorted as R-CE) loss down-weights the hard positive interactions, which is defined as:
%
%************************ formula 4 **************%
%
\begin{equation}\small
% \nonumber
\begin{aligned}
& \mathcal{L}_{\textit{R-CE}}(u, i) =  \omega(u, i) \mathcal{L}_{\textit{CE}}(u, i), 
\end{aligned}
\end{equation}
%
%************************ formula 4 **************%
%
where $\omega(u, i)$ is a weight function that adjusts the contribution of an interaction to the training objective. To properly down-weight the hard interactions, the weight function $\omega(u, i)$ is expected to have the following properties:
1) it dynamically adjusts the weights of interactions during training;
2) the function will reduce the influence of a hard interaction to be weaker than an easy interaction;
and 3) the extent of weight reduction can be easily adjusted to fit different models and datasets.

\begin{figure}[tb]
\setlength{\abovecaptionskip}{0cm}
\setlength{\belowcaptionskip}{0cm}
\centering 
\hspace{-0.2in}
    \subfigure[R-CE loss for the observed positive interactions. The contributions of large-loss interactions are greatly reduced.]{
    % \vspace{-0.1in}
    \label{fig5:subfig:a}
    \includegraphics[width=1.5in]{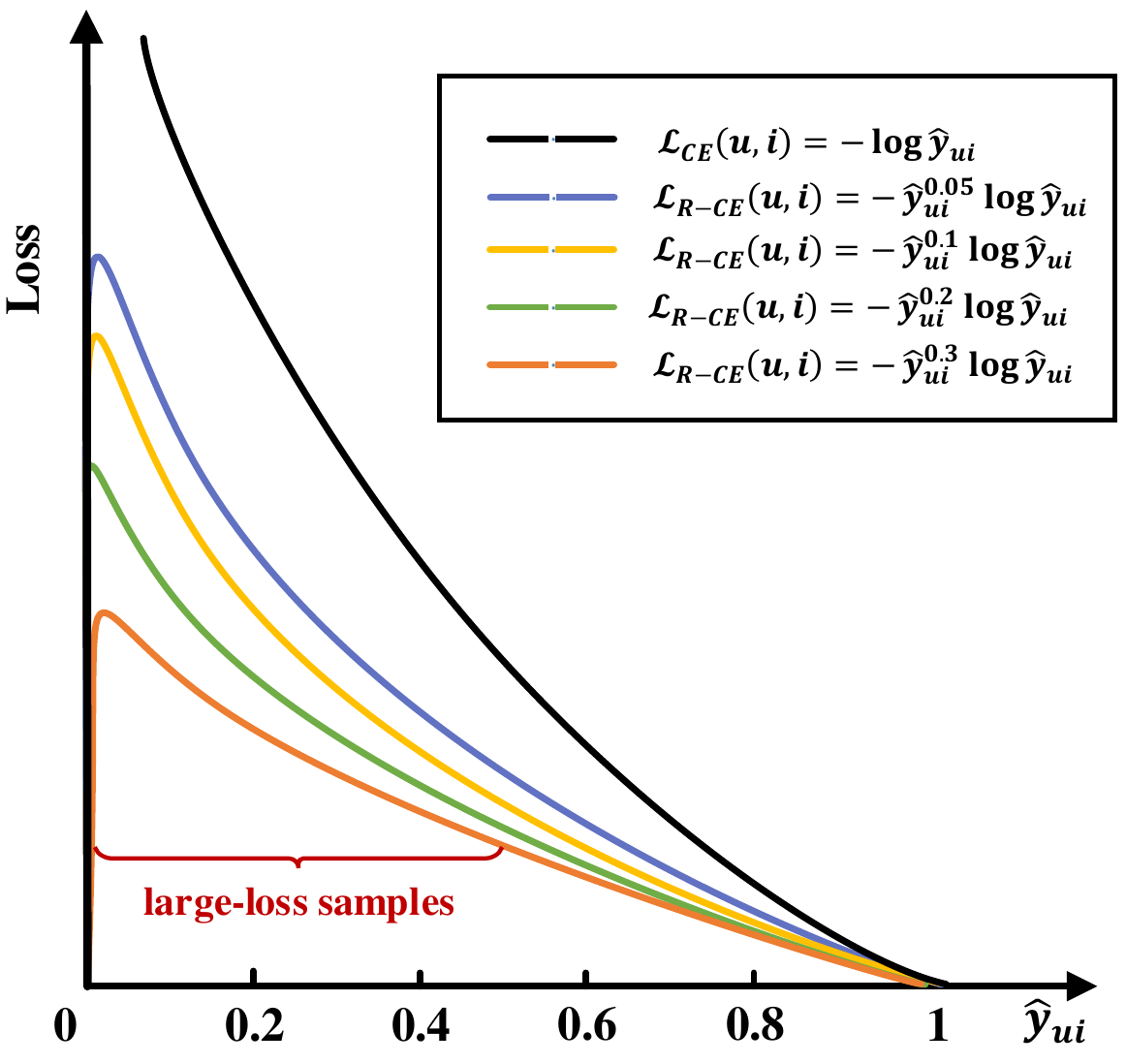}} 
    \hspace{0.05in}
    \subfigure[The weight function with different parameters $\beta$, where $\beta$ controls the weight difference between hard and easy interactions.]{
    % \vspace{-0.2in}
    \label{fig5:subfig:b} 
    \includegraphics[width=1.8in]{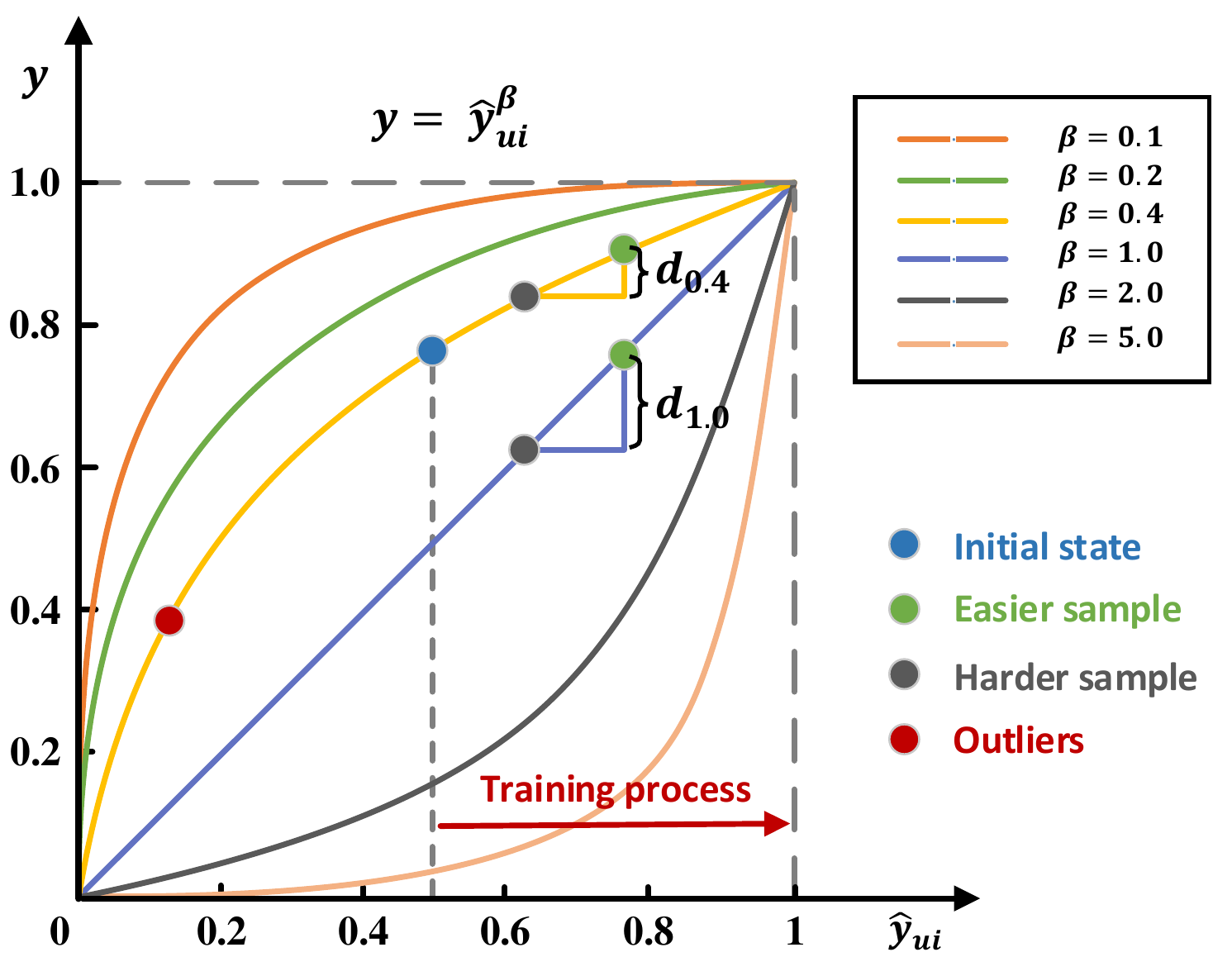}} 
  \caption{Illustration and analysis of R-CE loss.} 
    \label{fig:R-CE}
\end{figure}

Inspired by the success of Focal Loss~\cite{lin2017focal}, we estimate $\omega(u, i)$ according to the prediction score $\hat{y}_{ui}$. The prediction score is within $[0, 1]$ whereas the value of CE loss is in $[0, +\infty]$. And thus the prediction score is more accountable for further computation. Note that the smaller prediction score on the positive interaction leads to larger CE loss. 
Formally, we define the weight function as:
%
%************************ formula 5 **************%
%
\begin{equation}\small
 \omega(u, i) = \hat{y}_{ui}^{\beta},
\end{equation}
%
%************************ formula 5 **************%
%
where $\beta \in [0, +\infty]$ is a hyper-parameter to control the range of weights. 
From Figure \ref{fig5:subfig:a}, we can see that R-CE loss equipped with the proposed weight function can significantly reduce the loss of hard interactions (\eg $\hat{y}_{ui} \ll 0.5$) as compared to the original CE loss. Furthermore, the proposed weight function satisfies the aforementioned requirements:

\begin{itemize}[leftmargin=*]
    \item It generates dynamic weights during training since $\hat{y}_{ui}$ changes along the training process.
    \item The interactions with extremely large CE loss (\eg the ``outlier'' in Figure \ref{fig5:subfig:b}) will be assigned with very small weights because $\hat{y}_{ui}$ is close to 0. Therefore, the influence of such large-loss interactions is largely reduced. 
    In addition, as shown in Figure \ref{fig5:subfig:b}, harder interactions always have smaller weights because the function $f(\hat{y}_{ui})$ monotonically increases when $\hat{y}_{ui} \in [0, 1]$ and $\beta \in [0, +\infty]$. As such, it can avoid that false-positive interactions dominate the optimization during training \cite{Yang2018robust}.
    \item The hyper-parameter $\beta$ dynamically controls the gap between the weights of hard and easy interactions. By observing the examples in Figure \ref{fig5:subfig:b}, we can find that: 1) the gap between the weights of easy nd hard interactions becomes larger as $\beta$ increases (\eg $d_{0.4} < d_{1.0}$); and 2) the R-CE loss will degrade to a standard CE loss if $\beta$ is set as 0. 
\end{itemize}

In order to prevent negative interactions with large loss values from dominating the optimization, we revise the weight function as:
%
%************************ formula 5 **************%
%
\begin{equation}
\omega(u, i) = 
\begin{cases}
\hat{y}_{ui}^{\beta}, & \bar{y}_{ui}=1, \\
(1-\hat{y}_{ui})^{\beta}, & \text{otherwise.}
\end{cases}
\end{equation}
%
%************************ formula 5 **************%
%
Indeed, it may provide a possible solution to alleviate the impact of false-negative interactions, which is left for future exploration.

\subsubsection{\textbf{In-depth Analysis}}

Due to depending totally on recommenders to identify false-positive interactions, the reliability of ADT might be questionable. Actually, many existing work~\cite{lu2018MentorNet, han2018co} has pointed out the connection between the large loss and noisy interactions, and explained the underlying causality: the ``memorization'' effect of deep models. That is, deep models will first learn easy and clean patterns in the initial training phase, and then gradually memorize all interactions, including noisy ones. As such, the loss of deep models in the early stage can help to filter out noisy interactions. We discuss the memorization effect of recommenders by experiments in Section \ref{sec:observations} and \ref{sec:memorization}. 

Another concern is that discarding hard interactions would limit the model's learning ability since some hard interactions may be more informative than the easy ones. Indeed, as discussed in the prior studies on curriculum learning~\cite{Bengio2009Curriculum}, hard interactions in the noisy data probably confuse the model rather than help it to establish the right decision surface. As such, they may induce poor generalization. It's actually a trade-off between denoising and learning. In ADT, the $\epsilon(\cdot)$ of T-CE loss and $\beta$ of R-CE loss are to control the balance.

\section{Related Work}
% This work aims to denoise implicit feedback for recommenders, which is highly related to the negative experience identification, incorporating various feedback, and the robustness of recommenders.

%
%************************ Figure 6 **************%
%

\vspace{0cm}
\paragraph{\textbf{Negative Experience Identification}}
To reduce the gap between implicit feedback and the actual user preference, many researchers have paid attention to identify negative experiences in implicit signals \cite{fox2005Evaluating, Kim2014Modeling, jiang2020aspect}. Prior work usually collects the various users' feedback (\textit{e.g.}, dwell time \cite{Kim2014Modeling}, gaze patterns \cite{Zhao2016Gaze}, and skip \cite{fox2005Evaluating}) and the item characteristics \cite{Lu2018Between} to predict the user's satisfaction. 
% Lu \textit{et al.} \cite{Lu2018Between} predicted users' actual preference in news recommendation based on various user behaviors, news quality, and the interaction context. 
However, these methods need additional feedback and extensive manual label work, \eg users have to tell if they are satisfied for each interaction. Besides, the quantification of item quality is non-trivial \cite{Lu2018Between}, which largely relies on the manually feature design  and the labeling of domain experts \cite{Lu2018Between}. The unaffordable labor cost hinders the practical usage of these methods, especially in the scenarios with constantly changing items.

\vspace{-0.2cm}
\paragraph{\textbf{Incorporating Various Feedback}}

To alleviate the impact of false-positive interactions, previous approaches \cite{liu2010understanding, Yang2012Exploiting} also consider incorporating more feedback (\textit{e.g.}, dwell time~\cite{Yi2014Beyond}, skip~\cite{Wen2019Leveraging}, and adding to favorites) into training directly. For instance, Wen \textit{et al.} \cite{Wen2019Leveraging} proposed to train the recommender using three kinds of items: ``click-complete'', ``click-skip'', and ``non-click'' ones. The last two kinds of items are both treated as negative samples but with different weights. 
However, additional feedback might be unavailable in complex scenarios. 
For example, we cannot acquire dwell time and skip patterns after users buy products or watch movies in a cinema. Most users even don't give any informative feedback after clicks. In an orthogonal direction, this work explores denoising implicit feedback without additional data during training.

\vspace{-0.2cm}
\paragraph{\textbf{Robustness of Recommender Systems}}
Robustness of deep learning has become increasingly important \cite{feng2019graph, han2018co}. 
As to the robustness of recommenders, Gunawardana \textit{et al.} \cite{gunawardana2015evaluating} defined it as ``the stability of the recommendation in the presence of fake information''. Prior work \cite{Lam2004Shilling, Shriver2019Evaluating} has tried to evaluate the robustness of recommender systems under various attack methods, such as shilling attacks \cite{Lam2004Shilling} and fuzzing attacks \cite{Shriver2019Evaluating}. To build more robust recommender systems, some auto-encoder based models \cite{Florian2015Collaborative, wu2016collaborative, Liang2018Variational} introduce the denoising techniques. 
These approaches (\eg CDAE~\cite{wu2016collaborative}) first corrupt the interactions of user by random noises, and then try to reconstruct the original one with auto-encoders. 
However, these methods focus on heuristic attacks or random noises, and ignore the natural false-positive interactions in data. This work highlights the negative impact of natural noisy interactions, and improve the robustness against them.

\section{Experiment}

\paragraph{Dataset}\label{section5-1}
To evaluate the effectiveness of the proposed ADT on recommender training, we conducted experiments on three publicly accessible datasets: Adressa, Amazon-book, and Yelp. 
\begin{itemize}[leftmargin=*]
    \item \textbf{Adressa:} This is a real-world news reading dataset from Adressavisen\footnote{\url{https://www.adressa.no/}}~\cite{Gulla2017the}. It includes user clicks over news and the dwell time for each click, where the clicks with dwell time < 10s are thought of as false-positive ones~\cite{Yi2014Beyond, Kim2014Modeling}.
    \item \textbf{Amazon-book:} It is from the Amazon-review datasets\footnote{\url{http://jmcauley.ucsd.edu/data/amazon/}}~\cite{he2016ups}. It covers users' purchases over books with rating scores. A rating score below 3 is regarded as a false-positive interaction.
    \item \textbf{Yelp:} It's an open recommendation dataset\footnote{\url{https://www.yelp.com/dataset/challenge}}, in which businesses in the catering industry (\eg restaurants and bars) are reviewed by users. Similar to Amazon-book, the rating scores below 3 are regarded as false-positive feedback.
\end{itemize}

% These datasets comprise the common implicit feedback: click, purchase, and consumption, which are suitable to explore the effectiveness of denoising implicit feedback although explicit feedback also exists in each interaction.
We followed former work~\cite{Cheng2019MMALFM, LightGCN2020he} to split the dataset into training, validation, and testing (see Table \ref{table2} for statistics). To evaluate the effectiveness of denoising implicit feedback, we kept all interactions, including the false-positive ones, in training and validation, and tested recommenders only on true-positive interactions. That is, the models are expected to recommend more satisfying items.

%
%************************ table 2 **************%
%
\begin{table}[t]\small
\setlength{\abovecaptionskip}{0.1cm}
\setlength{\belowcaptionskip}{0cm}
\caption{Statistics of the datasets. In particular, \#FP interactions refer to the number of false-positive interactions.}
\label{table2}
\centering
\setlength{\tabcolsep}{1.3mm}{
\begin{tabular}{l|l|l|l|l}
\hline
Dataset      & \#User & \#Item & \#Interaction & \#FP Interaction       \\ \hline
Adressa & 212,231  & 6,596   & 419,491        & 247,628            \\ \hline
Amazon-book  & 80,464  & 98,663  & 2,714,021      & 199,475             \\ \hline
Yelp         & 45,548  & 57,396  & 1,672,520      & 260,581             \\ \hline
\end{tabular}
}
% \vspace{-0.3cm}
\end{table}
%
%************************ table 2 **************%
%
%
%************************ table 3 **************%
%
\begin{table*}[t]
% \vspace{-0.2cm}
\setlength{\abovecaptionskip}{0.1cm}
\setlength{\belowcaptionskip}{0cm}
\caption{Overall performance of three testing recommenders trained with ADT strategies and normal training over three datasets. Note that Recall@K and NDCG@K are shorted as R@K and N@K to save space, respectively, and ``RI'' in the last column denotes the relative improvement of ADT over normal training on average. The best results are highlighted in bold.}
\label{table3}
\centering
\setlength{\tabcolsep}{1.3mm}{
\begin{tabular}{l|cccc|cccc|cccc|c}
\hline
\textbf{Dataset} & \multicolumn{4}{c|}{\textbf{Adressa}} & \multicolumn{4}{c|}{\textbf{Amazon-book}} & \multicolumn{4}{c|}{\textbf{Yelp}} &  \\ %\hline
\textbf{Metric} & \textbf{R@3} & \textbf{R@20} & \textbf{N@3} & \textbf{N@20} & \textbf{R@50} & \textbf{R@100} & \textbf{N@50} & \textbf{N@100} & \textbf{R@50} & \textbf{R@100} & \textbf{N@50} & \textbf{N@100} & \textbf{RI} \\ \hline \hline
GMF & 0.0880 & 0.2141 & 0.0780 & 0.1237 & 0.0609 & 0.0949 & 0.0256 & 0.0331 & 0.0840 & 0.1339 & 0.0352 & 0.0465 & -\\ %\hline
GMF+T-CE & \textbf{0.0892} & \textbf{0.2170} & \textbf{0.0790} & \textbf{0.1254} & \textbf{0.0707} & \textbf{0.1113} & \textbf{0.0292} & \textbf{0.0382} & \textbf{0.0871} & \textbf{0.1437} & 0.0359 & \textbf{0.0486} & 7.14\% \\ %\hline
GMF+R-CE  & 0.0891 & 0.2142 & 0.0765 & 0.1229 & 0.0682 & 0.1075 & 0.0275 & 0.0362 & 0.0861 & 0.1361 & \textbf{0.0366} & 0.0480 & 4.34\% \\ \hline
NeuMF & 0.1416 & 0.3159 & 0.1267 & 0.1886 & 0.0512 & 0.0829 & 0.0211 & 0.0282 & 0.0750 & 0.1226 & 0.0304 & 0.0411 & -\\ %\hline
NeuMF+T-CE & \textbf{0.1418} & 0.3106 & 0.1227 & 0.1840 &  \textbf{0.0725} &  \textbf{0.1158} &  \textbf{0.0289} &  \textbf{0.0385} & \textbf{0.0825} & \textbf{0.1396} & \textbf{0.0323} & \textbf{0.0451} & 15.62\% \\ %\hline
NeuMF+R-CE & {0.1414} & \textbf{0.3185} & {0.1266} & \textbf{0.1896} & {0.0628} & {0.1018} & {0.0248} & {0.0334} & 0.0788 & 0.1304 & 0.0320 & 0.0436 & 8.77\% \\ \hline
CDAE & 0.1394 & 0.3208 & 0.1168 & 0.1808 & 0.0989 & 0.1507 & 0.0414 & 0.0527 & 0.1112 & 0.1732 & 0.0471 & 0.0611 &  - \\ %\hline
CDAE+T-CE & \textbf{0.1406} & \textbf{0.3220} & 0.1176 & \textbf{0.1839} & \textbf{0.1088} & \textbf{0.1645} & \textbf{0.0454} & \textbf{0.0575} & \textbf{0.1165} & \textbf{0.1806} & \textbf{0.0504} & \textbf{0.0652} & 5.36\% \\ %\hline
CDAE+R-CE & 0.1388 & 0.3164 & \textbf{0.1200} & 0.1827 & 0.1022 & 0.1560 & 0.0424 & 0.0542 & 0.1161 & 0.1801 & 0.0488 & 0.0632 & 2.46\% \\ \hline
\end{tabular}
}
\end{table*}
%
%************************ table 3 **************%
%

% \vspace{-0.1cm}
\paragraph{Evaluation Protocols}\label{sec:protocol}
For each user in the testing set, we predicted the preference score over all the items except the positive ones used during training. Following existing studies \cite{He2017Neural, wang2019NGCF}, we reported the performance \wrt two widely used metrics: Recall@K and NDCG@K, where higher scores indicate better performance. For both metrics, we set K as 50 and 100 for Amazon-book and Yelp, while 3 and 20 for Adressa due to its much smaller item space.

% \vspace{-0.1cm}
\paragraph{Testing Recommenders}
To demonstrate the effectiveness of our proposed ADT strategy on denoising implicit feedback, we compared the performance of recommenders trained with T-CE or R-CE and normal training with standard CE. We selected two representative user-based neural CF models, GMF and NeuMF~\cite{He2017Neural}, and one item-based model, CDAE~\cite{wu2016collaborative}. Note that CDAE is also a representative model of robust recommender which can defend random noises within implicit feedback.

\begin{itemize}[leftmargin=*]
    \item \textbf{GMF \cite{He2017Neural}:} This is a generalized version of matrix factorization by replacing the inner product with the element-wise product and a linear neural layer as the interaction function.
    \item \textbf{NeuMF \cite{He2017Neural}:} NeuMF models the relationship between users and items by combining GMF and a Multi-Layer Perceptron (MLP). 
    \item \textbf{CDAE \cite{wu2016collaborative}:} CDAE corrupts the interactions with random noises, and then employs a MLP model to reconstruct the original input.
    % partly increasing its anti-noise capability. 
\end{itemize}
We only tested neural models and omit conventional ones such as MF and SVD++~\cite{Koren2008Factorization} due to their inferior performance~\cite{He2017Neural, wu2016collaborative}.

% \vspace{-0.1cm}
\paragraph{Parameter Settings}

For the three testing recommenders, we followed their default settings, and verified the effectiveness of our methods under the same conditions. For GMF and NeuMF, the factor numbers of users and items are both 32. As to CDAE, the hidden size of MLP is set as 200. In addition, Adam~\cite{kingma2014adam} is applied to optimize all the parameters with the learning rate initialized as 0.001 and he batch size set as 1,024. As to the ADT strategies, they have three hyper-parameters in total: $\alpha$ and $\epsilon_{max}$ in T-CE loss, and $\beta$ in R-CE loss. In detail, $\epsilon_{max}$ is searched in $\{0.05, 0.1, 0.2\}$ and $\beta$ is tuned in $\{0.05, 0.1, ..., 0.25, 0.5, 1.0\}$. As for $\alpha$, we controlled its range by adjusting the iteration number $\epsilon_{N}$ to the maximum drop rate $\epsilon_{max}$, and $\epsilon_{N}$ is adjusted in $\{1k, 5k, 10k, 20k, 30k\}$.

% \vspace{-0.1cm}
\subsection{Performance Comparison}
Table \ref{table3} summarizes the recommendation performance of the three testing models trained with standard CE, T-CE, or R-CE over three datasets. From Table \ref{table3}, we can observe:
\begin{itemize}[leftmargin=*]
    \item In all cases, both the T-CE loss and R-CE loss effectively improve the performance, \eg NeuMF+T-CE outperforms vanilla NeuMF by 15.62\% on average over three datasets. The significant performance gain indicates the better generalization ability of neural recommenders trained by T-CE loss and R-CE loss. It validates the effectiveness of adaptive denoising training, \ie discarding or down-weighting hard interactions during training. 
    \item By comparing the T-CE Loss and R-CE Loss, we found that the T-CE loss performs better in most cases. We postulate that the recommender still suffers from the false-positive interactions when it is trained with R-CE Loss, even though they have smaller weights and contribute little to the overall training loss. In addition, the superior performance of T-CE Loss might be attributed to the additional hyper-parameters in the dynamic threshold function which can be tuned more granularly. Further improvement might be achieved by a finer-grained user-specific or item-specific tuning of these parameters, which can be done automatically~\cite{lambdaopt}.
    \item Across the recommenders, NeuMF performs worse than GMF and CDAE, especially on Amazon-book and Yelp, which is criticized for the vulnerability to noisy interactions. Because our testing is only on the true-positive interactions, the inferior performance of NeuMF is reasonable since NeuMF with more parameters can fit more false-positive interactions during training.
    \item Both T-CE and R-CE achieve the biggest performance increase on NeuMF, which validates the effectiveness of ADT to prevent vulnerable models from the disturbance of noisy data. On the contrary, the improvement over CDAE is relatively small, showing that the design of defending random noise can also improve the robustness against false-positive interactions to some extent. Nevertheless, applying T-CE or R-CE still leads to performance gain, which further validates the rationality of denoising implicit feedback.
\end{itemize}

\noindent In the following, GMF is taken as an example to conduct thorough investigation for the consideration of computation cost.

\paragraph{Further Comparison against Using Additional Feedback}

To avoid the detrimental effect of false-positive interactions, a popular idea is to incorporate the additional user feedback for training although they are usually sparse. Existing work either adopts the additional feedback by multi-task learning \cite{Gao2019LearningTR}, or leverages it to identify the true-positive interactions \cite{Wen2019Leveraging, Lu2018Between}. In this work, we introduced two classical models for comparison: Neural Multi-Task Recommendation (NMTR)~\cite{Gao2019LearningTR} and Negative feedback Re-weighting (NR)~\cite{Wen2019Leveraging}. In particular, NMTR consider both click and the additional feedback with multi-task learning. NR uses the addition feedback (\ie dwell time or rating) to identify true-positive interactions and re-weights the false-positive and non-interacted ones. We applied NMTR and NR on the testing recommenders and reported the results of GMF in Table \ref{table4}. From Table \ref{table4}, we can find that: 1) NMTR and NR achieve slightly better performance than GMF, which validates the effectiveness of additional feedback; and 2) the results of NMTR and NR are inferior to that of T-CE and R-CE. This might be attributed to the sparsity of the additional feedback. Indeed, the clicks with satisfaction is much fewer than the total number of clicks, and thus NR will lose extensive positive training interactions. 
% Besides, not all clicks without labeled user satisfaction indicate users' dislikes because many users seldom give explicit feedback even if they are satisfied. Therefore, treating them as negative interactions will hurt the performance, which is also found by the experiments in \cite{ding2019sampler}. 

\begin{table}[]
% \vspace{-0.25cm}
\setlength{\abovecaptionskip}{0.1cm}
\caption{Performance \wrt GMF on Amazon-book.}
\label{table4}
\centering
\setlength{\tabcolsep}{1.3mm}{
\begin{tabular}{l|cccc}
\hline
\textbf{Metric} & \textbf{R@50} & \textbf{R@100} & \textbf{N@50} & \textbf{N@100} \\ \hline \hline
{GMF} & 0.0609 & 0.0949 & 0.0256 & 0.0331 \\ \hline
{GMF+T-CE} & \textbf{0.0707} & \textbf{0.1113} & \textbf{0.0292} & \textbf{0.0382} \\ %\hline
{GMF+R-CE} & 0.0682 & 0.1075 & 0.0275 & 0.0362 \\ \hline
{GMF+NMTR} & 0.0616 & 0.0967 & 0.0254 & 0.0332 \\ %\hline
{GMF+NR} & 0.0615 & 0.0958 & 0.0254 & 0.0331  \\ \hline
\end{tabular}
}
\vspace{0.2cm}
\end{table}

% \vspace{-0.15cm}
\paragraph{Performance Comparison w.r.t. Interaction Sparsity}

\begin{figure}
\vspace{-0.3cm}
\setlength{\abovecaptionskip}{0.cm}
  \centering 
  \hspace{-0.5in}
  \subfigure{
    \includegraphics[width=1.8in]{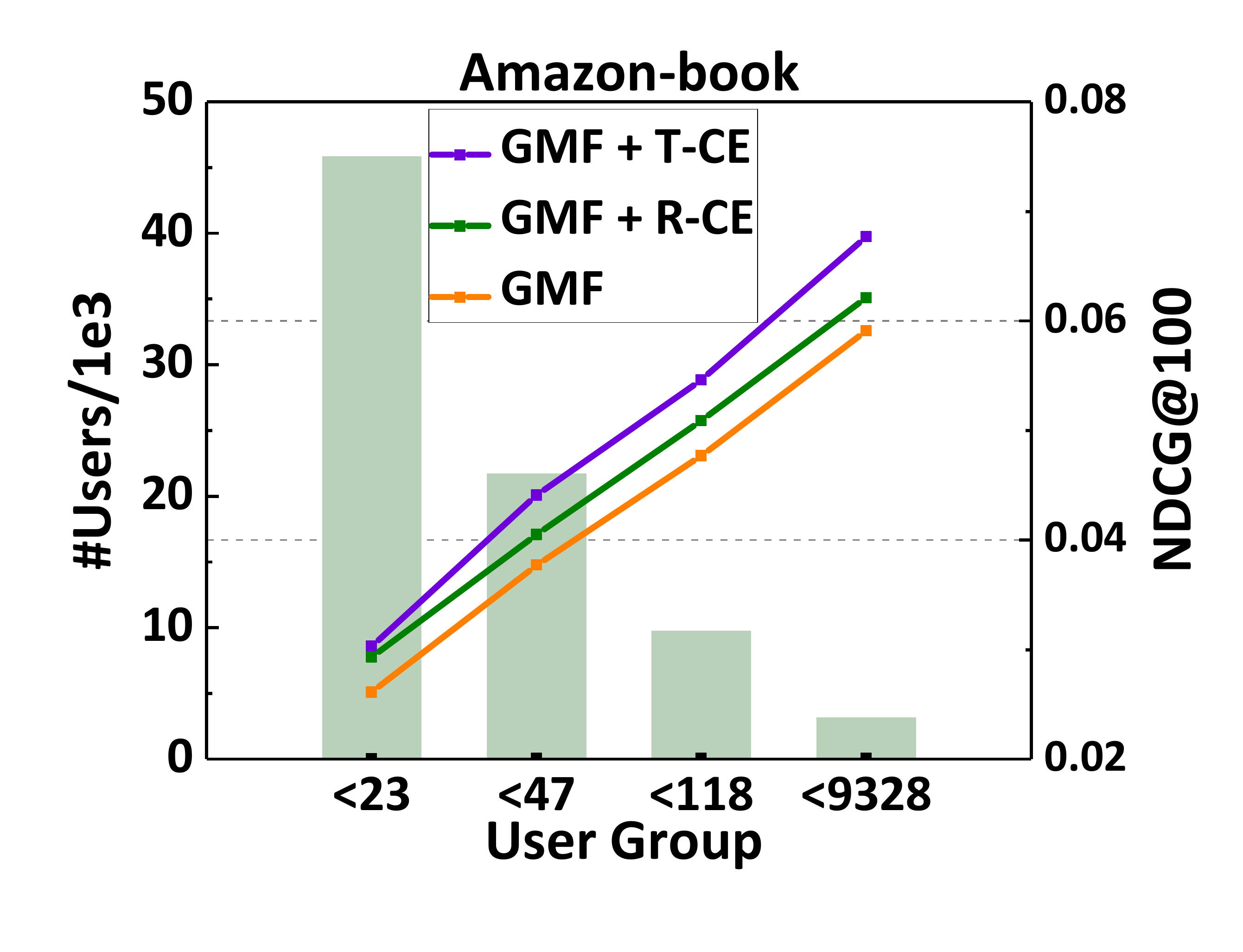}} 
  \hspace{-0.2in}
  \subfigure{
    \includegraphics[width=1.8in]{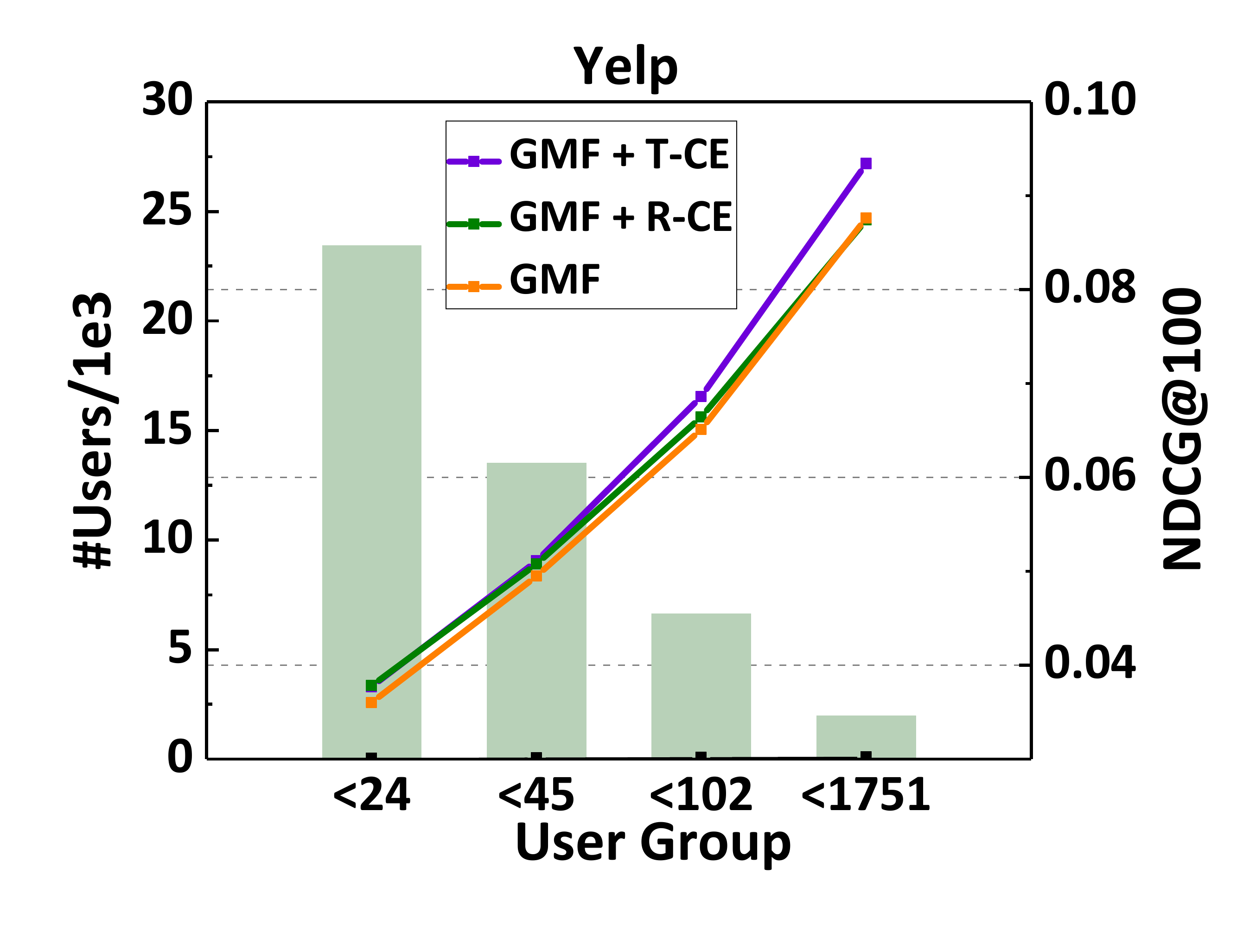}} 
  \hspace{-0.5in} 
%   \vspace{-0.1in}
  \caption{Performance comparison of GMF over user groups with different sparsity levels. The histograms represent the user number and the lines denote the performance.} 
  \label{fig:usergroup}
\end{figure}

Since ADT prunes many interactions during training, we explored whether ADT hurts the preference learning of inactive users because their interacted items are sparse. Following the former studies~\cite{wang2019NGCF}, we split testing users into four groups according to the interaction number of each user where each group has the same number of interactions. Figure \ref{fig:usergroup} shows the group-wise performance comparison where we can observe that the proposed ADT strategies achieve stable performance gain over normal training in all cases. It validates that ADT is also effective for the inactive users. 

\begin{figure}
\vspace{-0.4cm}
\setlength{\abovecaptionskip}{0.1cm}
  \centering 
  \hspace{-0.6in}
  \subfigure[Normal Training]{
    \includegraphics[width=1.42in]{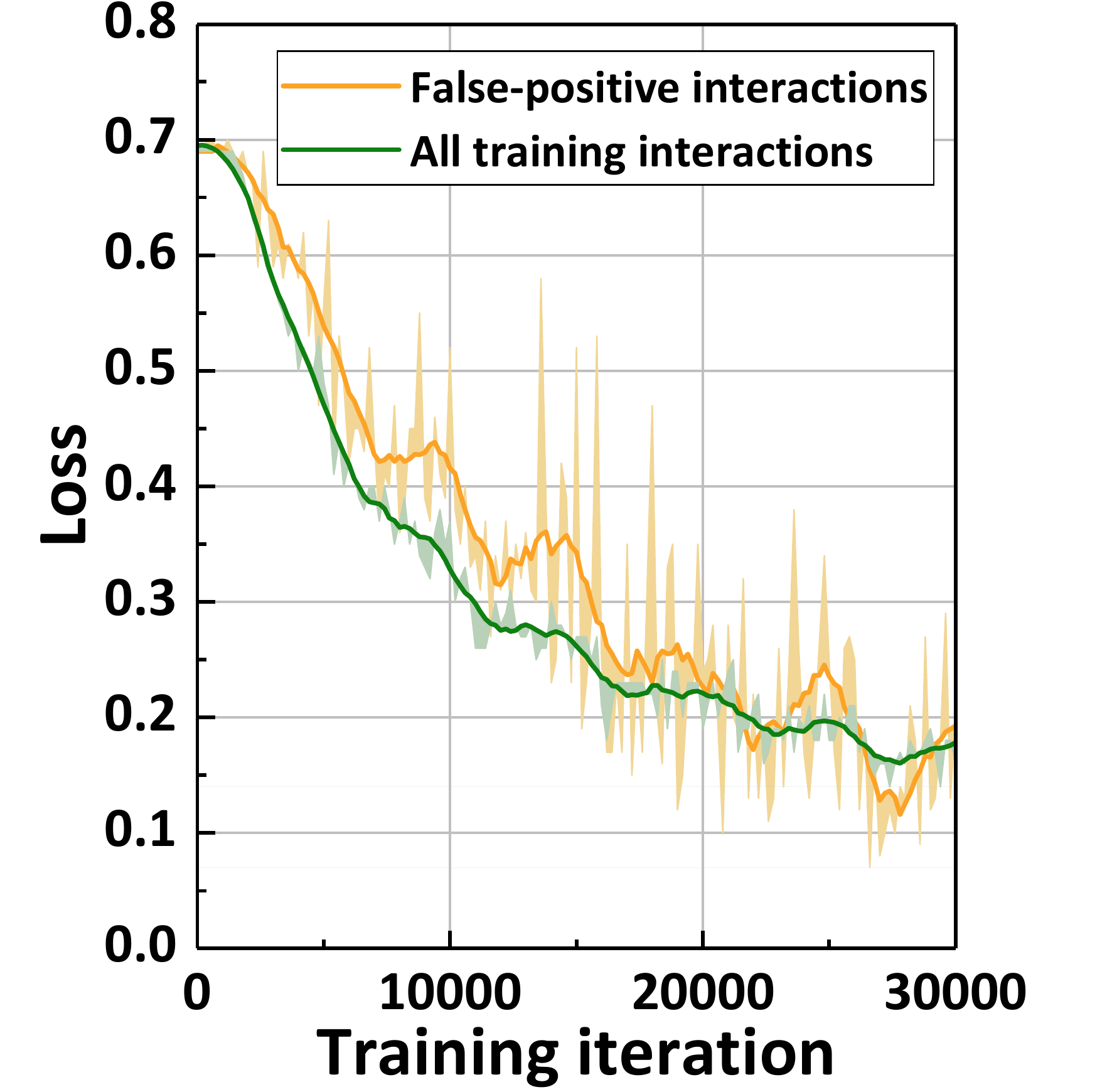}} 
  \hspace{-0.28in}
  \subfigure[Truncated Loss]{
    \includegraphics[width=1.42in]{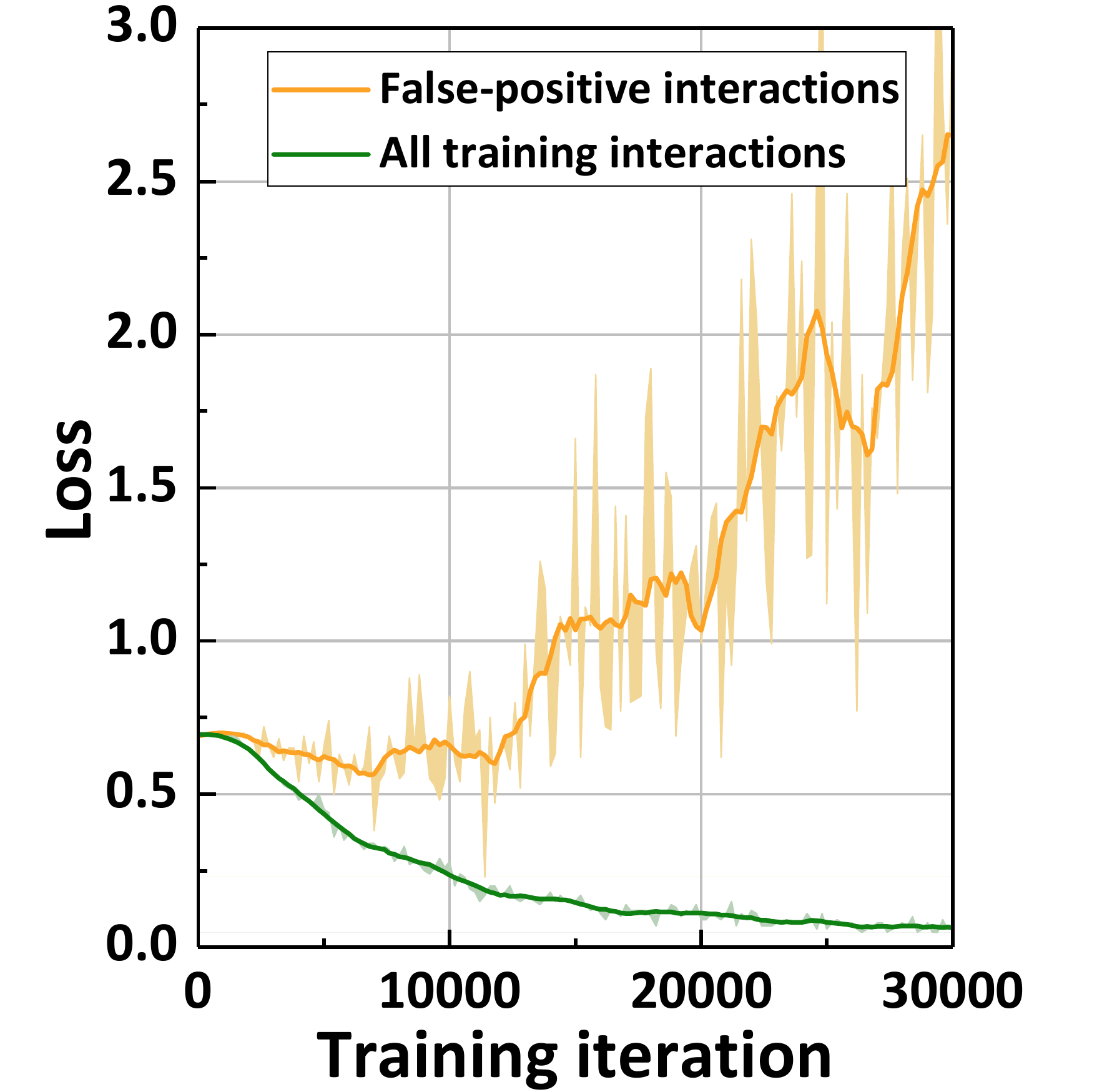}} 
  \hspace{-0.28in}
  \subfigure[Reweighted Loss]{
    \includegraphics[width=1.42in]{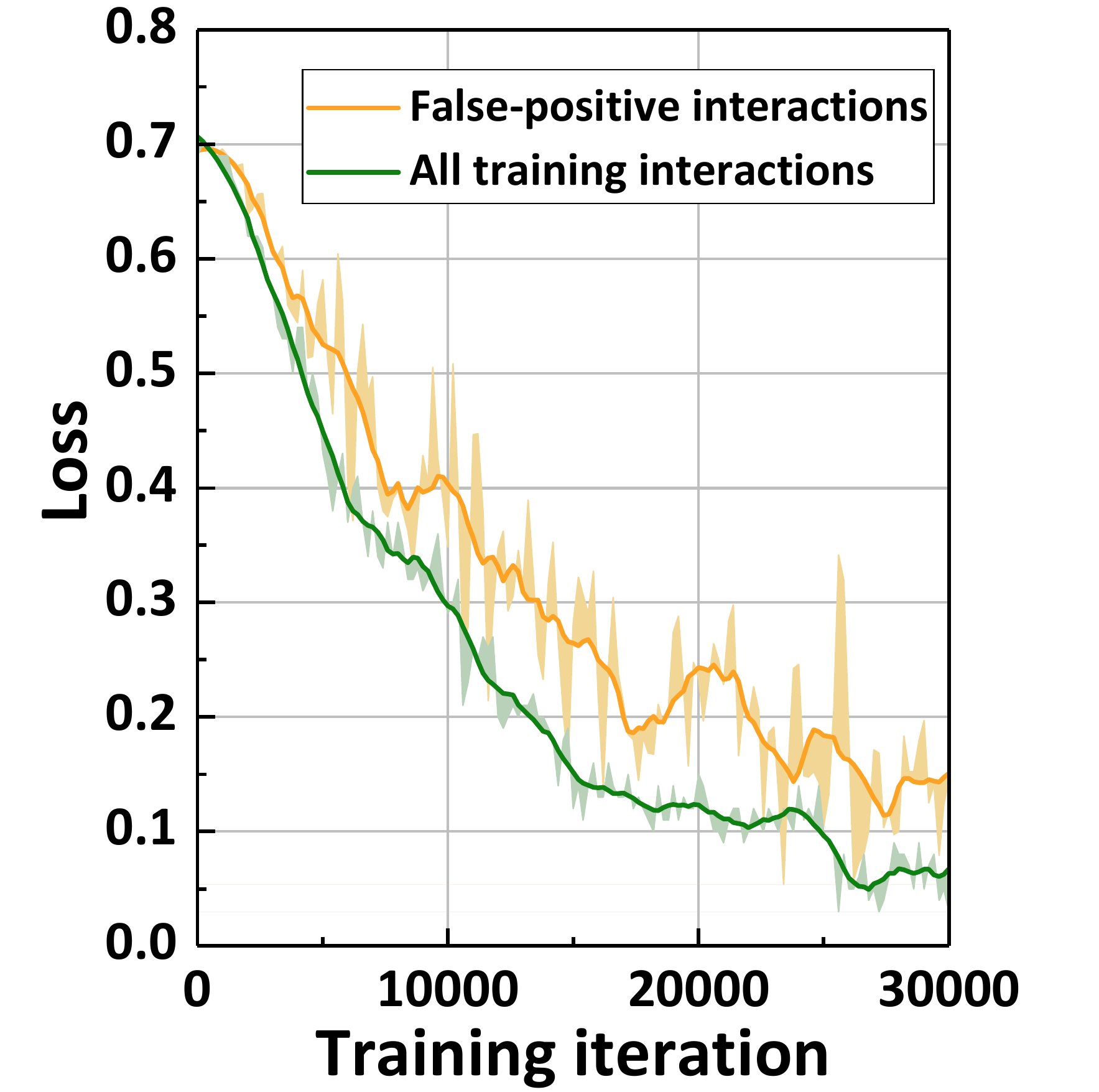}} 
  \hspace{-0.6in} 
%   \vspace{-0.1in}
  \caption{Loss of GMF (a), GMF+T-CE (b) and GMF+R-CE (c).} 
  \label{fig:lossTrend}
  \vspace{0.2cm}
\end{figure}

\begin{figure*}[ht]
% \vspace{-0.2cm}
\setlength{\abovecaptionskip}{-0.5cm}
\setlength{\belowcaptionskip}{0cm}
  \centering 
  \hspace{-0.6in}
  \subfigure{
    \includegraphics[width=1.3in]{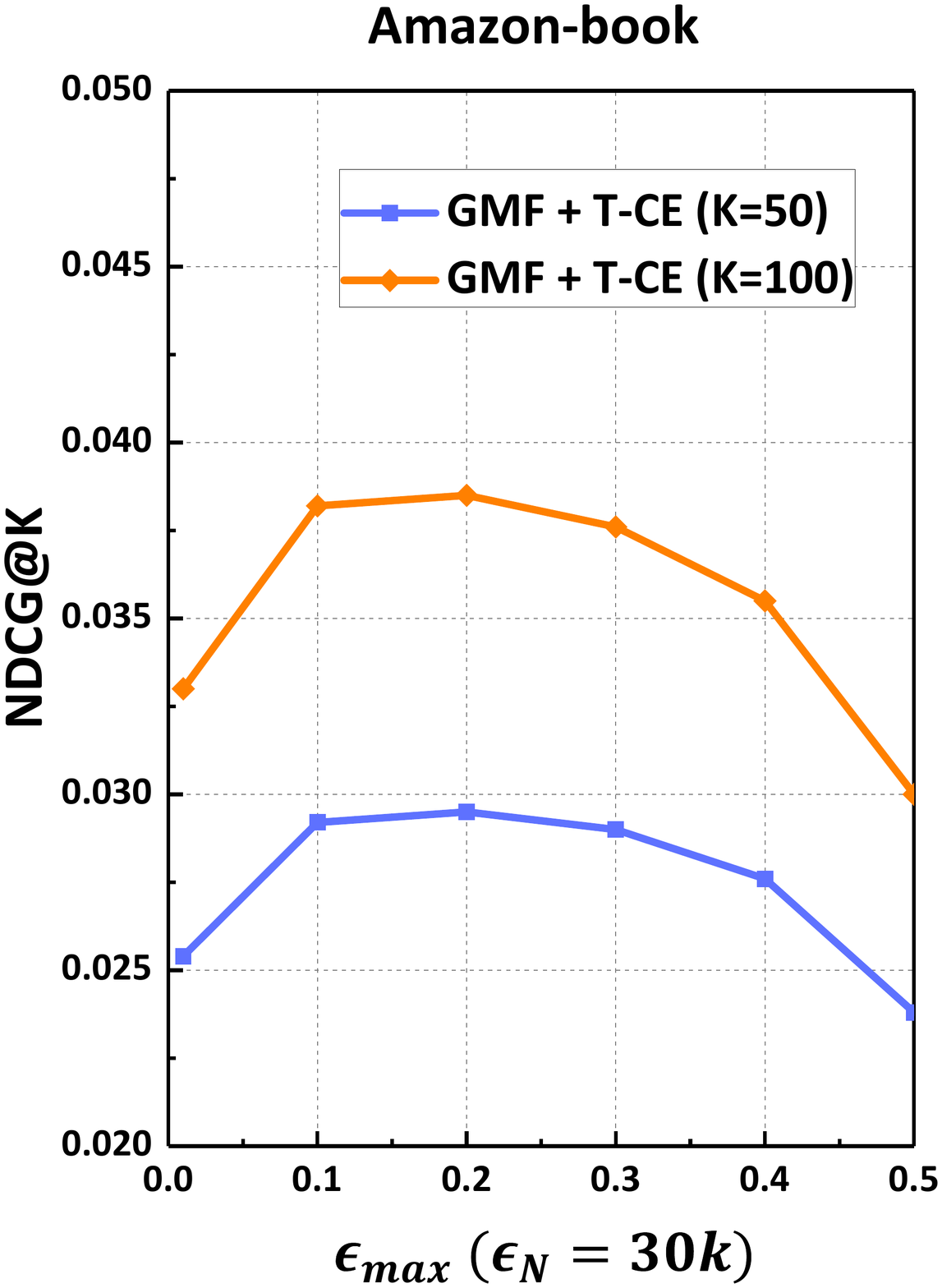}} 
  \hspace{-0.25in}
  \subfigure{
    \includegraphics[width=1.3in]{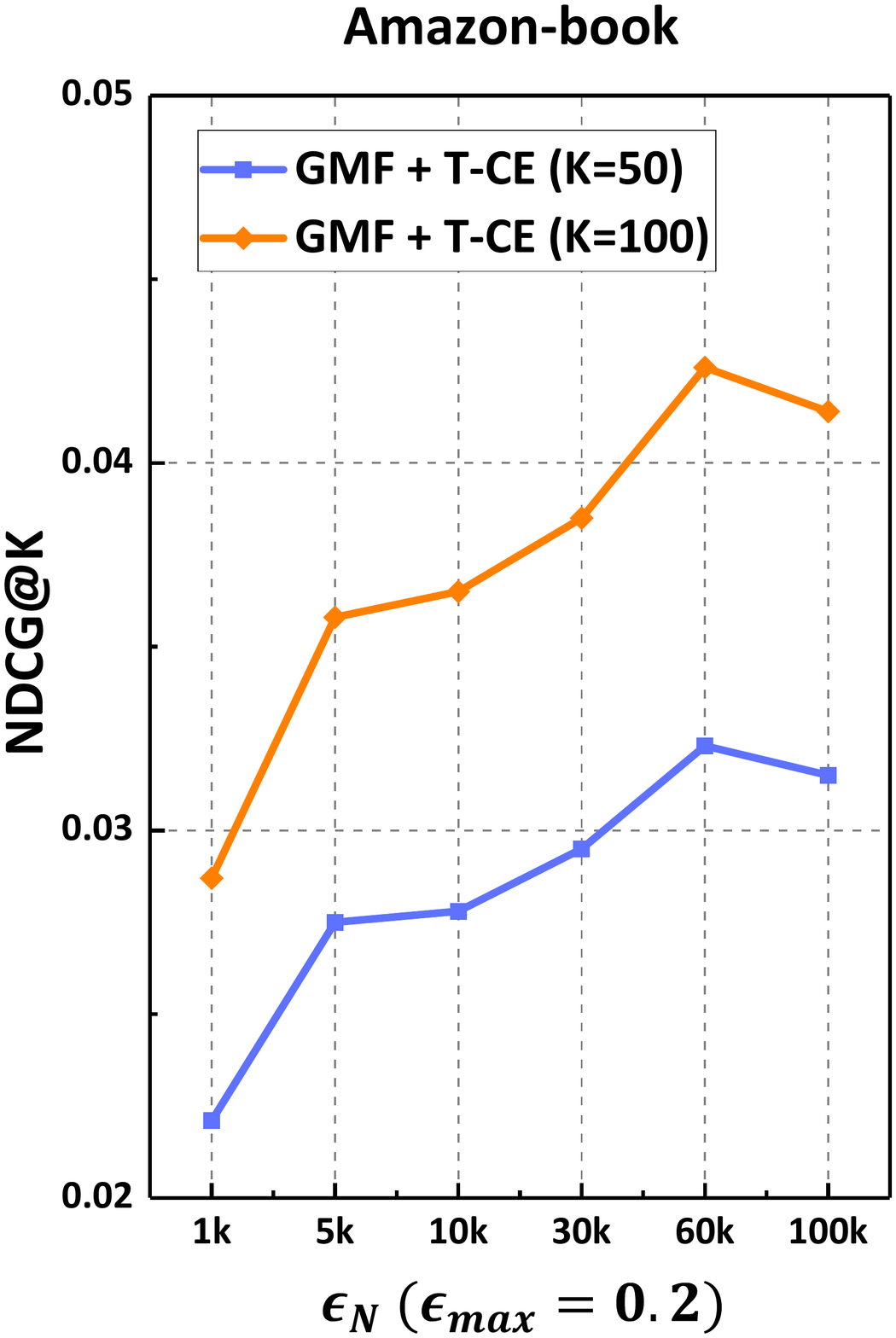}} 
  \hspace{-0.25in}
  \subfigure{
    \includegraphics[width=1.3in]{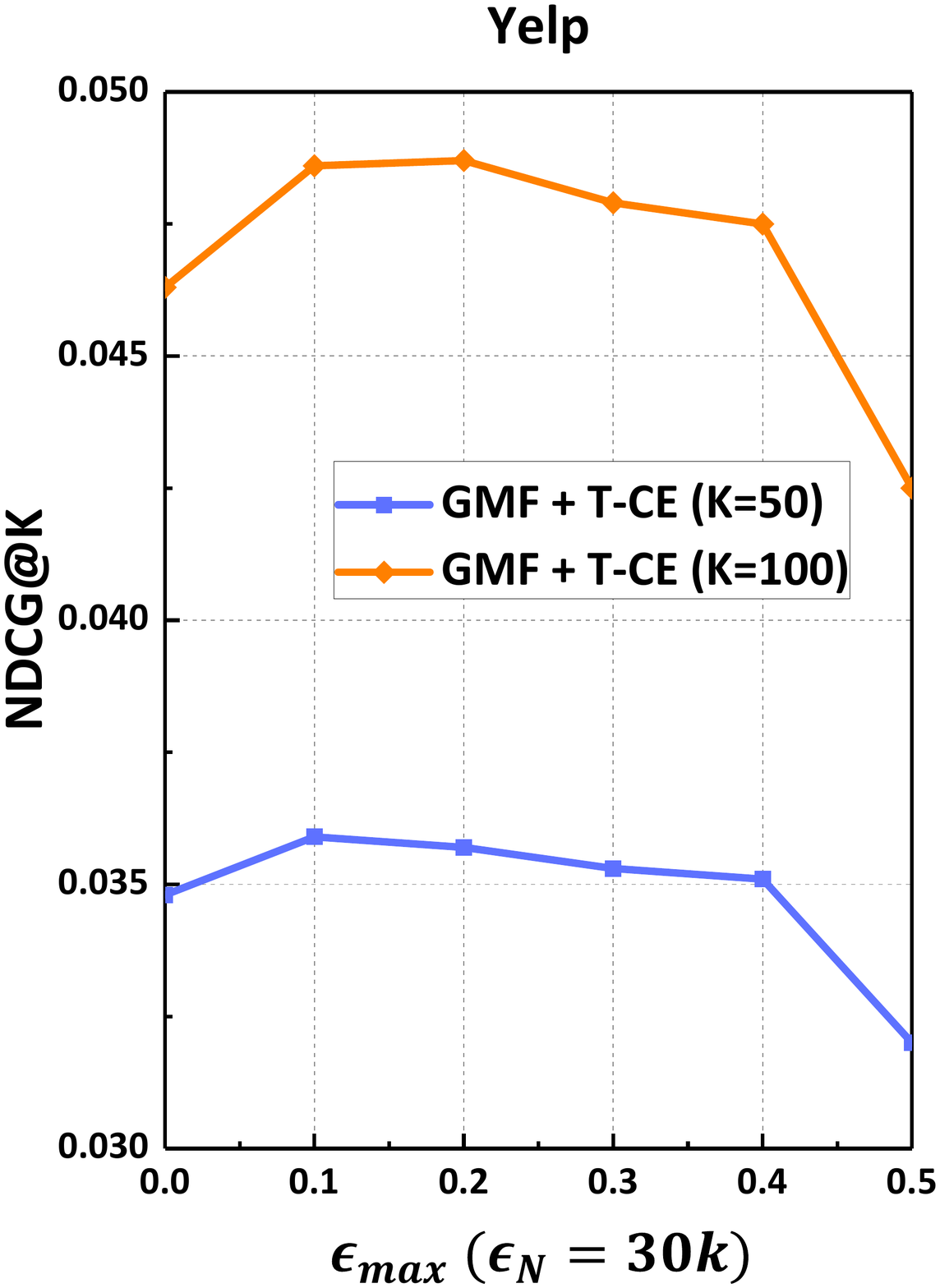}} 
  \hspace{-0.25in}
  \subfigure{
    \includegraphics[width=1.3in]{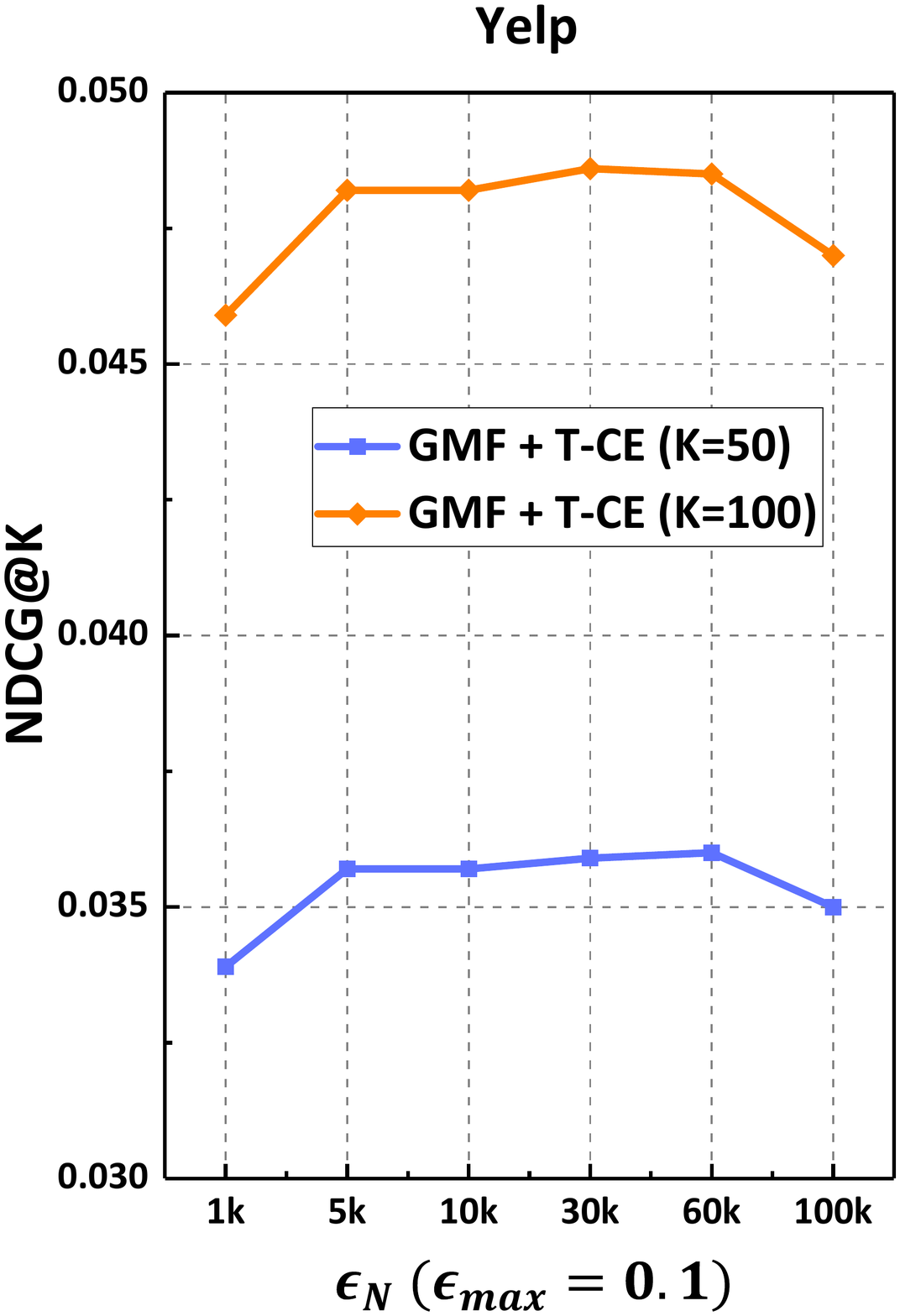}} 
  \hspace{-0.25in}
  \subfigure{
    \includegraphics[width=1.3in]{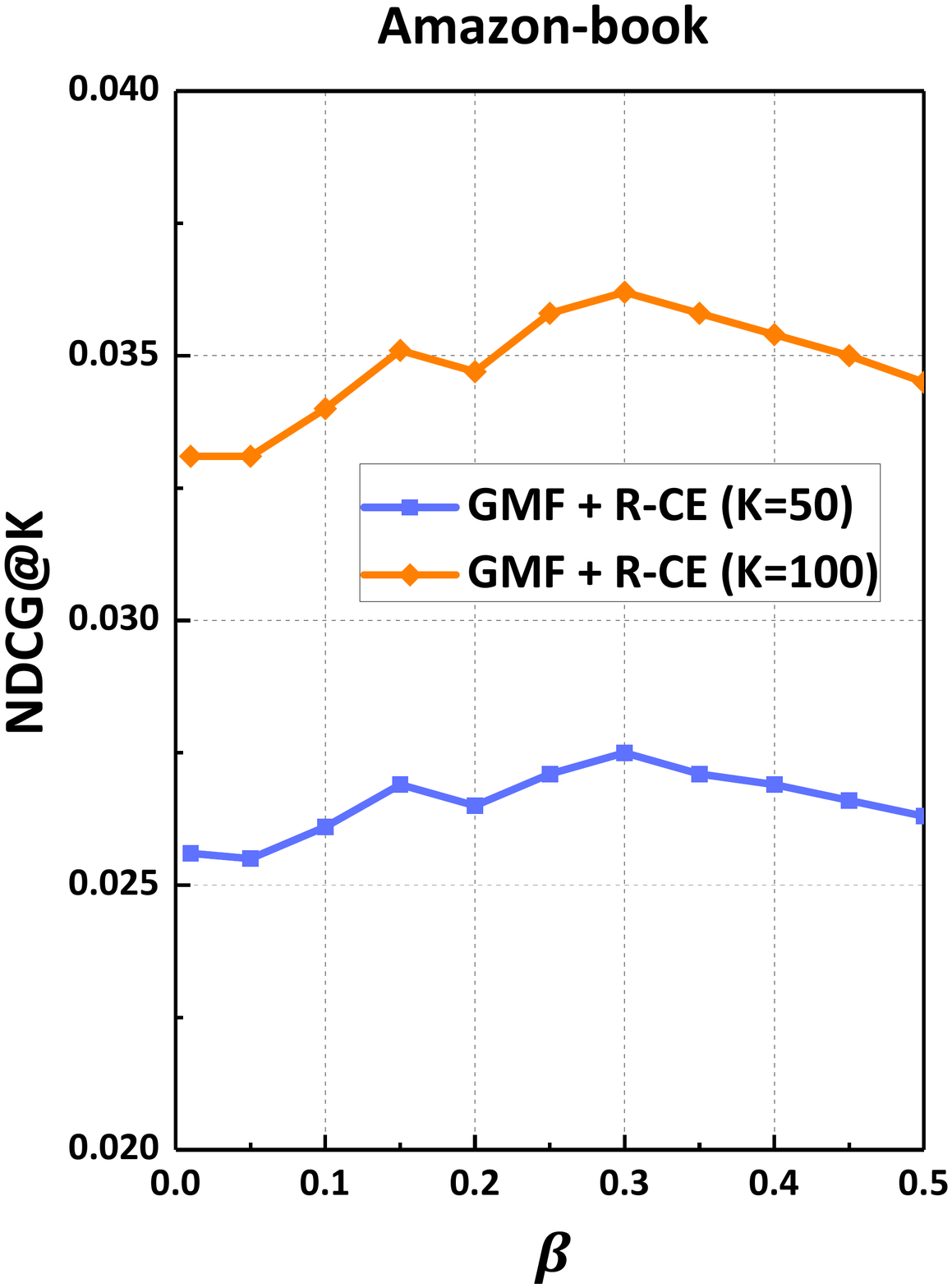}} 
  \hspace{-0.25in}
  \subfigure{
    \includegraphics[width=1.3in]{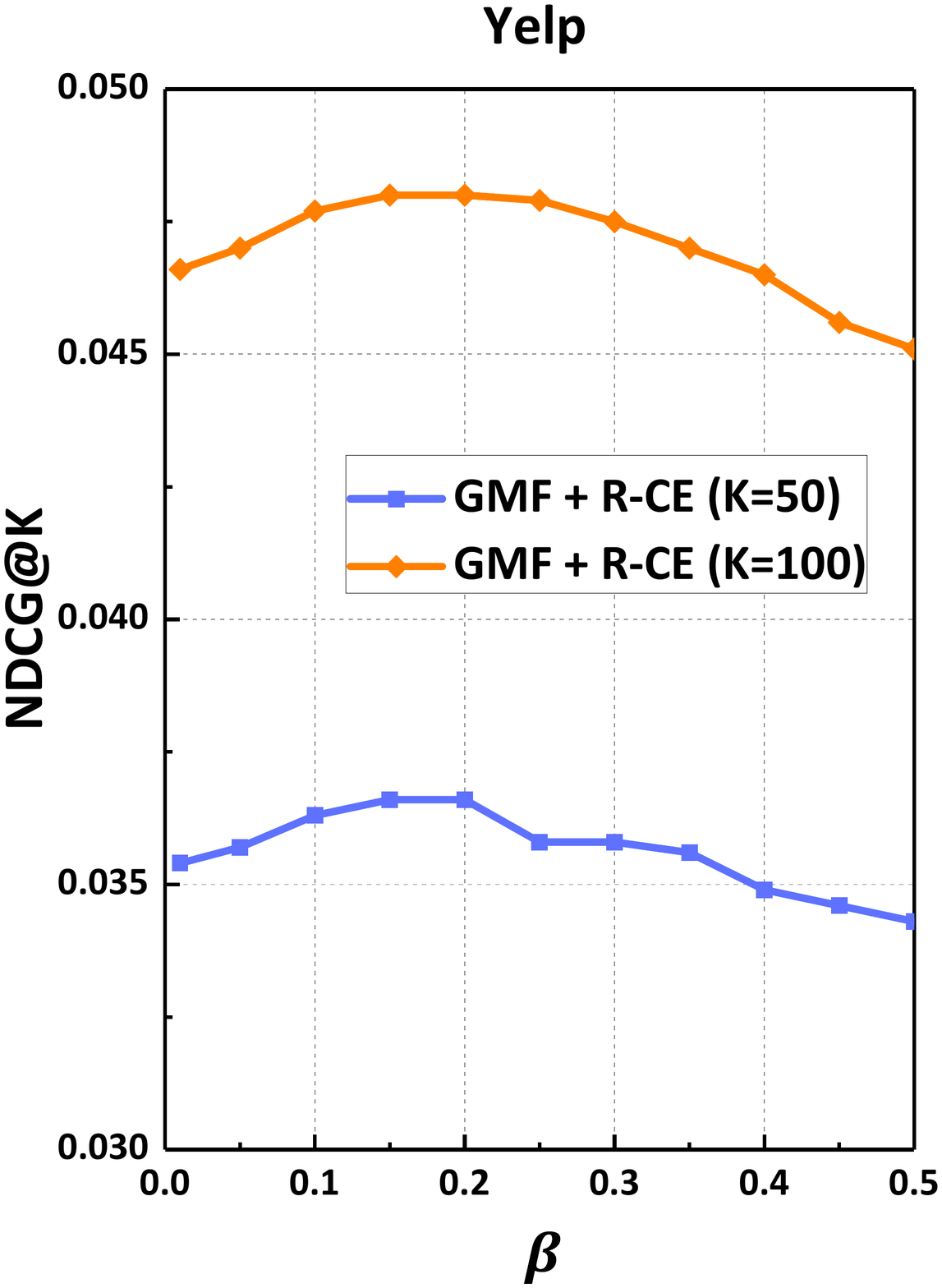}} 
  \hspace{-0.6in} 
%   \vspace{-0.1in}
  \caption{Performance comparison of GMF trained with ADT on Yelp and Amazon-book \wrt different hyper-parameters.} 
  \label{fig:parameter}
\end{figure*}

\begin{figure}[t]
\vspace{-0.2cm}
\setlength{\abovecaptionskip}{0cm}
\setlength{\belowcaptionskip}{-0cm}
  \centering 
  \hspace{-0.5in}
  \subfigure{
    \includegraphics[width=1.8in]{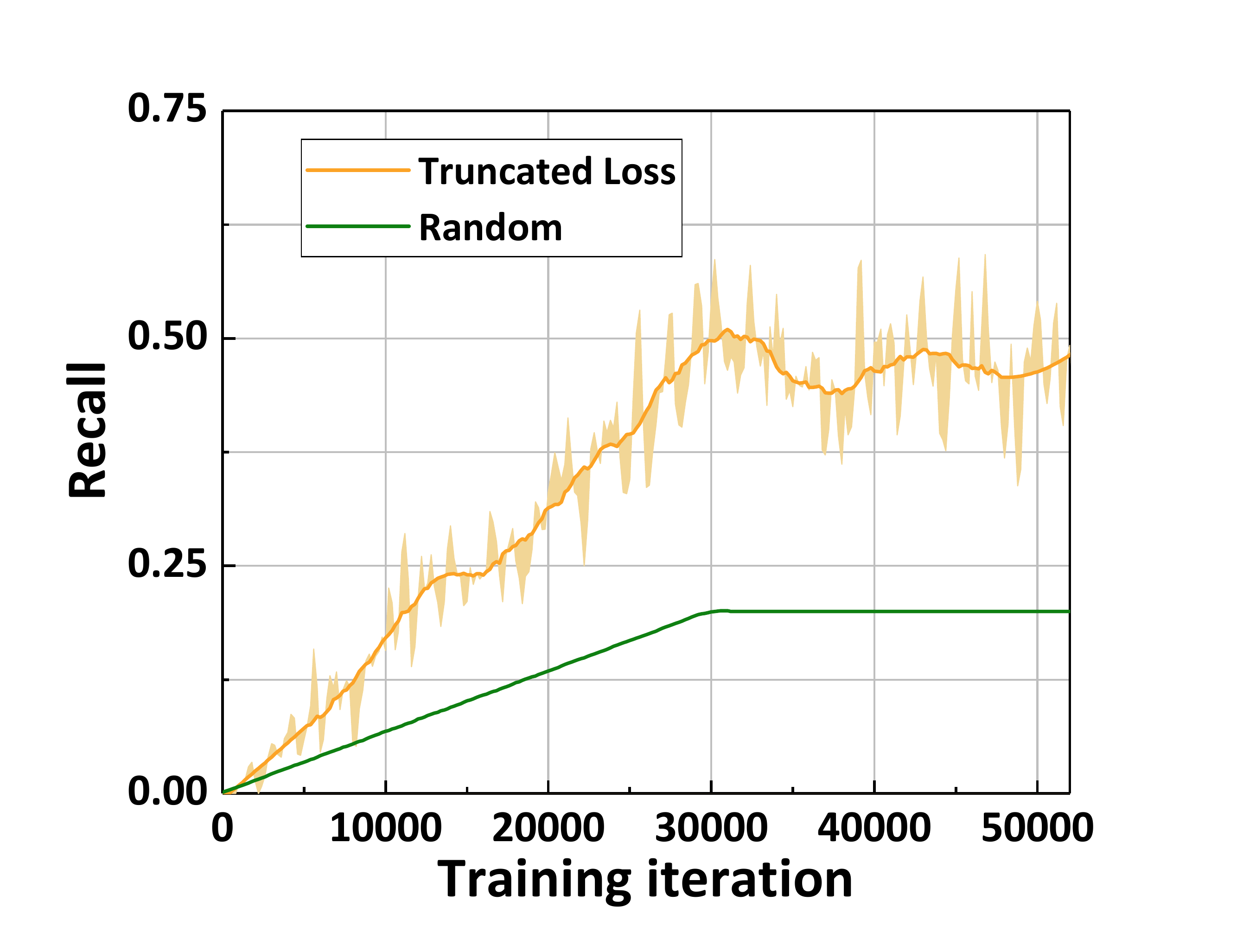}} 
  \hspace{-0.35in}
  \subfigure{
    \includegraphics[width=1.8in]{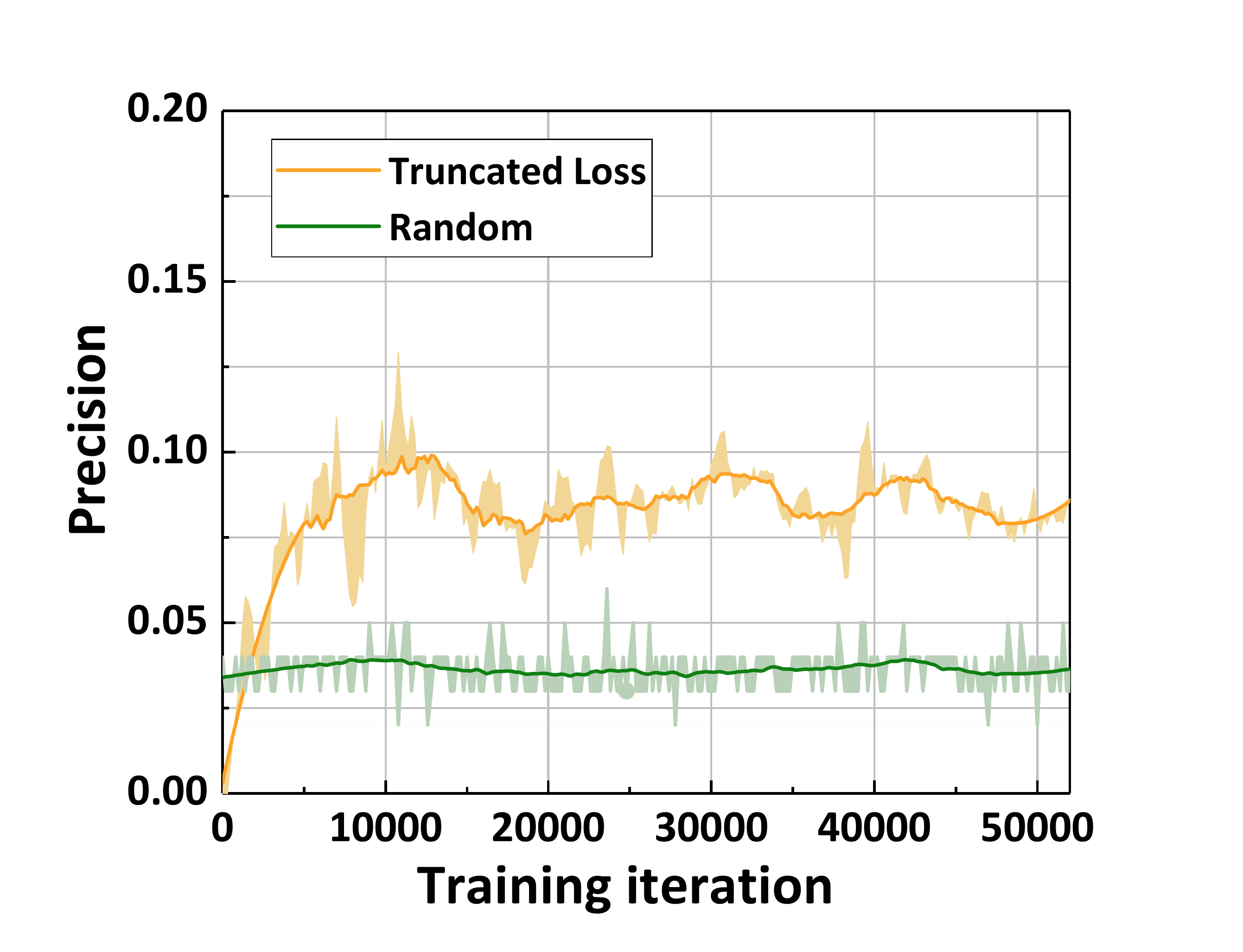}}
  \hspace{-0.5in} 
%   \vspace{-0.1in}
  \caption{Recall and precision of false-positive interactions over GMF trained with the Truncated Loss on Amazon-book.} 
  \label{fig:recall}
  \vspace{0.2cm}
\end{figure}

\subsection{In-depth Analysis}
\subsubsection{Memorization of False-positive Interactions}\label{sec:memorization}

Recall that false-positive interactions are memorized by recommenders eventually under normal training. We then investigated whether false-positive interactions are also fitted well by the recommenders trained with ADT strategies. 
We depicted the CE loss of false-positive interactions during the training procedure with the real training loss as reference.

From Figure \ref{fig:lossTrend}(a), we can find the CE loss values of false-positive interactions eventually become similar to other interactions, indicating that GMF fits false-positive interactions well at last. On the contrary, as shown in Figure \ref{fig:lossTrend}(b), by applying T-CE, the CE loss values of false-positive interactions increase while the overall training loss stably decreases step by step. The increased loss indicates that the recommender parameters are not optimized over the false-positive interactions, validating the capability of T-CE to identify and discard such interactions. As to R-CE (Figure \ref{fig:lossTrend}(c)), the CE loss of false-positive interactions also shows a decreasing trend, showing that the recommender still fits such interactions. However, their loss values are still larger than the real training loss, indicating assigning false-positive interactions with smaller weights by R-CE loss is effective. It prevents the model from fitting them quickly. Therefore, we can conclude that both paradigms reduce the effect of false-positive interactions on recommender training, which can explain their improvement over normal training.

% \vspace{-0.15cm}
\subsubsection{Study of Truncated Loss}\label{sec:studyTCE}
Since the Truncated Loss achieves promising performance, we studied how accurate it identifies and discards false-positive interactions. We first defined \textit{recall} to represent what percentage of false-positive interactions in the training data are discarded, and \textit{precision} as the ratio of discarded false-positive interactions to all discarded interactions. Figure \ref{fig:recall} visualizes the changes of the recall and precision along the training process. We take random discarding as a reference, where the recall of random discarding equals to the drop rate $\epsilon(T)$ during training and the precision is the proportion of noisy interactions within the training batch at each iteration. 

From Figure \ref{fig:recall}, we observed that: 1) the Truncated Loss discards nearly half of false-positive interactions after the drop rate keeps stable, greatly reducing the impact of noisy interactions; and 2) 
% the precision of Truncated Loss is about twice as large as that of random discarding. It demonstrates that the Truncated Loss effectively utilizes the distill signals of false-positive interactions and weakens their contributions to the model training. In spite of this, we can find that 
a key limitation of the Truncated Loss is the low precision, \eg only 10\% precision in Figure \ref{fig:recall}, which implies that it inevitably discards many clean interactions. However, it's worth pruning noisy interactions at the cost of losing many clean interactions. Besides, how to further improve the precision so as to decrease the loss of clean interactions is a promising research direction in the future.

% \vspace{-0.15cm}
\subsubsection{Hyper-parameter Sensitivity}\label{sec:hyperparameter}

The proposed ADT strategies introduces additional three hyper-parameters to adjust the dynamic threshold function and the weight function in two paradigms. In particular, $\epsilon_{max}$ and $\epsilon_{N}$ are used to control the drop rate in the Truncated Loss, and $\beta$ adjusts the weight function in the Reweighted Loss. We studied how the hyper-parameters affect the performance. 
% Only the results of GMF trained with ADT strategies on Amazon-book and Yelp are reported in Figure \ref{fig:parameter} due to space limitation. Other methods over three datasets have similar patterns. 
From Figure \ref{fig:parameter}, we can find that: 1) the recommender trained with the T-CE loss performs better when $\epsilon_{max} \in [0.1, 0.3]$. If $\epsilon_{max}$ exceeds 0.4, the performance drops significantly because a large proportion of interactions are discarded. Therefore, the upper bound $\epsilon_{max}$ should be restricted. 2) The recommender is relatively sensitive to $\epsilon_{N}$, especially on Amazon-book, and the performance still increases when $\epsilon_{N}$ > 30k. Nevertheless, a limitation of T-CE loss is the big search space of hyper-parameters. 3) The adjustment of $\beta$ in the Reweighted Loss is consistent over different datasets, and the best results happen when $\beta$ ranges from 0.15 to 0.3. These observations provide insights on how to tune the hyper-parameters of ADT if it's applied to other recommenders and datasets.

% \vspace{cm}
\section{Conclusion and Future Work}

In this work, we explored to denoise implicit feedback for recommender training. We found the negative effects of noisy implicit feedback, and proposed Adaptive Denoising Training strategies to reduce their impact. In particular, this work contributes two paradigms to formulate the loss functions: Truncated Loss and Reweighted Loss. 
% The former truncates the loss values of noisy samples to 0 with a dynamic threshold function; the latter reweighs all interactions adaptively during training. 
Both paradigms are general and can be applied to different recommendation loss functions, neural recommenders, and optimizers. In this work, we applied the two paradigms on the widely used binary cross-entropy loss and conduct extensive experiments over three recommenders on three datasets, showing that the paradigms effectively reduce the disturbance of noisy implicit feedback.

This work takes the first step to denoise implicit feedback for recommendation without using additional feedback for training, and points to some new research directions. Specifically, it is interesting to explore how the proposed two paradigms perform on other loss functions, such as Square Loss~\cite{Rosasco2004are}, Hinge Loss~\cite{Rosasco2004are} and BPR Loss~\cite{rendle2009bpr}. Besides, how to further improve the precision of the paradigms is worth studying. Lastly, our Adaptive Denoising Training is not specific to the recommendation task, and it can be widely used to denoise implicit interactions in other domains, such as Web search and question answering.

\begin{acks}
This research is supported by the National Research Foundation, Singapore under its International Research Centres in Singapore Funding Initiative, and the National Natural Science Foundation of China (61972372, U19A2079). Any opinions, findings and conclusions or recommendations expressed in this material are those of the author(s) and do not reflect the views of National Research Foundation, Singapore.
\end{acks}

\newpage
{
\tiny
\bibliographystyle{ACM-Reference-Format}
\balance
\bibliography{bibtex}
}
\end{document}